\documentclass[twocolumn,
nofootinbib,
 amsmath,amssymb,
 aps,
 prd,
 floatfix,
 superscriptaddress, 
]{revtex4}
\usepackage{scalerel}
\usepackage{tabularray}
\usepackage{graphicx}% Include figure files
\usepackage{dcolumn}% Align table columns on decimal point
\usepackage{bm}% bold math
\usepackage[normalem]{ulem}
\usepackage[center]{caption}
\usepackage{xcolor}

\usepackage{hyperref}
\usepackage{cleveref}
\usepackage{gensymb}
\begin{document}

%\preprint{APS/123-QED}

\title{Discriminating scalar ultralight dark matter from quasi-monochromatic gravitational waves in LISA}% Force line breaks with \\

\author{Jordan Gu\'e}\email{jgue@ifae.es}
%\email{jordan.gue@obspm.fr}
\affiliation{Institut de F\'isica d’Altes Energies (IFAE), The Barcelona Institute of Science and Technology, Campus UAB, 08193 Bellaterra (Barcelona), Spain}
\affiliation{%
 LTE, Observatoire de Paris, Université PSL, Sorbonne Université, Université de Lille, LNE, CNRS, 61 Avenue de l’Observatoire, 75014 Paris, France\\
}%
\author{Peter Wolf}
\affiliation{%
 LTE, Observatoire de Paris, Université PSL, Sorbonne Université, Université de Lille, LNE, CNRS, 61 Avenue de l’Observatoire, 75014 Paris, France\\
}%
\author{Aur\'elien Hees}
\affiliation{%
 LTE, Observatoire de Paris, Université PSL, Sorbonne Université, Université de Lille, LNE, CNRS, 61 Avenue de l’Observatoire, 75014 Paris, France\\
}%

\begin{abstract}

A scalar ultralight dark matter (ULDM) candidate would induce oscillatory motion of freely falling test masses via its coupling to Standard Model fields. Such oscillations would create an observable Doppler shift of light exchanged between the test masses, and in particular would be visible in space-based gravitational waves (GW) detectors, such as LISA. While this kind of detection has been proposed multiple times in the recent years, we numerically investigate if it is possible to extract a scalar ULDM signal in a space-based GW detector, and in particular how to differentiate such a signal from a GW signal. Using one year of realistic orbits for the LISA spacecrafts and Bayesian methods, we find that LISA will indeed be able to discriminate between the two signals.
\end{abstract}

\maketitle

\section{\label{sec:Intro}Introduction}

The microscopic nature of dark matter (DM) remains to this day one of the biggest puzzles of fundamental physics \cite{Bertone18}. While its gravitational impact on surrounding visible objects is well known (see e.g. \cite{Cirelli24} for a review), the fundamental interactions of DM with the Standard Model particles are still to be discovered. Over the whole zoo of possible candidates, ultralight dark matter (ULDM), i.e. DM with mass well below the eV, makes an appealing solution to the DM problem, in particular it can address some challenges of the $\Lambda$CDM model at galactic scales, see e.g. \cite{Ferreira21}. At cosmological scales, ULDM can be described as a classical bosonic field which oscillates at its Compton frequency $\omega$, when $\omega \gg H$, the Hubble constant. In the Milky Way's galactocentric frame, assumed to be the DM rest frame (on average), the DM particles acquire a non-zero velocity dispersion which implies that the DM wave is not purely monochromatic but presents a characteristic frequency dispersion $\Delta \omega \sim \omega/10^6$ \cite{Derevianko18,foster:2018aa}. 

Among those ULDM candidates, scalar fields, like dilaton or axionlike-particle, referred to as the axion in the following, induce spacetime variations of fundamental constants of Nature, like the fine structure constant or the electron mass through their couplings with various sectors of the Standard Model (SM) \cite{damour:2010zr,Kim:2022ype}. As a consequence, the rest mass and the transition frequency of test masses or atoms are also dependent on the spacetime position, therefore such oscillations can be probed accurately using atomic clocks or tests of the universality of free fall (UFF) like atom interferometry \cite{Hees16,hees18,Badurina22,Gue:2024onx}.

It has recently been proposed to use gravitational wave (GW) detectors to probe such couplings by measuring the phase shift of the laser light exchanged between the test masses \cite{Morisaki19,Yu23}.
The space-based GW observatory LISA (Laser Interferometer Space Antenna), as a joint mission between ESA (European Space Agency) and NASA (North American Space Agency), has recently been adopted \cite{Colpi24} and is expected to be launched in 2035. It consists of three spacecrafts (S/C) in heliocentric orbit, separated by 2.5 million kilometers armlength. The six individual optical interferometers will serve to detect GW in the $10^{-4}-1$ Hz frequency band, and in particular, LISA is expected to detect GW emitted by galactic binaries (GBs) \cite{Timpano06,Toonen17}. In this case, the two massive bodies are far from the coalescence, such that the timescale of their inspiral phase is much larger than the mission duration, or in other words the signal is quasi-monochromatic. GW act on LISA arms through Doppler shifts of light exchanged between the S/C, induced by the oscillation of the physical distance between the test masses. 

DM also induces quasi-monochromatic oscillations of the physical distance between the test masses via the oscillations of their mass and thus their momentum. Therefore, both the DM and GW signals produce similar signals in Fourier domain, in the form of a quasi-monochromatic peak at the wave frequency. Then, a natural question arises:  will space-based GW detector such as LISA be able to disentangle a scalar ULDM signal from a GW one?
This is the question we aim to answer in this paper. Using Bayesian inference on mock scalar ULDM and monochromatic GW signals in one year of data generated using realistic orbits of the S/C \cite{martens:2021aa} based on \textit{LISAOrbits} \cite{Bayle22}, we evaluate if one is able to differentiate a DM signal from a GW one. Our main conclusion is that there is no degeneracy between the DM and GW signals. In other words, this means that Bayesian model selection allows to clearly identify an injected scalar ULDM signal within the data with respect to a GW signal. On the other hand, Bayesian model selection clearly shows that a scalar ULDM model cannot explain an injected monochromatic GW signal, such that the monochromatic GW model is preferred over the ULDM one. Therefore, we conclude that there is no substantial degeneracies between the two dataset as seen by LISA, and that the expected sensitivity curves on the dilatonic couplings computed in \cite{Yu23} are valid. 

The paper is organised as follows. In Sec.~\ref{sec:theory}, we start by computing the acceleration induced by scalar DM candidates. Then, in Sec.~\ref{sec:LISA_signatures}, we show how these accelerations produce an observable signal in space-based GW detectors, and we compare it with the signal induced by monochromatic GW. In Sec.~\ref{sec:data_analysis}, we dig into the main part of the paper, namely we describe our Bayesian analysis on the distinction between ULDM and GW signals in LISA, from the methodology to the results. In Sec.~\ref{sec:results}, we compute the sensitivity of LISA to ULDM couplings, and we discuss our results in Sec.~\ref{sec:discussion}.

\section{Test mass oscillating acceleration induced by scalar dark matter}\label{sec:theory}
In this section, we briefly review some known results regarding the acceleration induced by scalar dark matter on test masses (see e.g. \cite{damour:2010zr,hees18,Gue:2024onx} for more details).

\subsection{Scalar field linearly coupled to Standard Model}\label{sec:dilaton_lin}

We will be interested in the coupling between a dimensionless dilaton scalar field $\phi$ and the Standard Model, with the corresponding low-energy effective linear interaction Lagrangian \cite{damour:2010zr}
\begin{align}
\mathcal{L}_\mathrm{int} &=\phi \left(\frac{d_e}{4\mu_0}F_{\mu\nu}F^{\mu\nu} - \frac{d_g\beta_3}{2g_3}G^a_{\mu\nu}G^{a\mu\nu} - \right.\,\label{dilaton_lagrangian}\\
&\left.\sum_{i=e,u,d}(d_{m_i} +\gamma_{m_i}d_g)m_i \bar{\psi_i}\psi_i\right) \, \nonumber ,
\end{align}
where $F^{\mu \nu}, G^{a\mu \nu}$ represent the electromagnetic and gluonic strength tensors respectively, $\mu_0$ is the vacuum magnetic permeability, $g_3, \beta_3$ are the dimensionless QCD coupling constant and QCD beta function for the running of $g_3$, $m_i$ the mass of the fermions fields in units of energy, with spinors $\psi_i$ and $\gamma_{m_i}$ the anomalous dimension giving the energy running of the masses of the QCD coupled fermions. The $d_i$ represent the dimensionless coupling constants between the scalar dilaton and the different matter fields. 

These interactions between the SM particles and the dilaton lead to variation of several fundamental constants of Nature, namely the QCD energy scale $\Lambda_3$, the electromagnetic fine structure constant $\alpha$, the electron mass $m_e$ and the light quark masses $m_i$, with $i=\{u,d\}$ as \cite{damour:2010zr}
\begin{subequations}\label{const_variations}
\begin{align}
\alpha(\phi) &= \left(1+d_e \phi\right)\alpha \, \\
\Lambda_3(\phi) &= \left(1+d_g\phi\right)\Lambda_3 \, , \\
m_e(\phi) &= \left(1+d_{m_e}\phi\right)m_e \, , \\
m_i(\Lambda_3)(\phi) &= \left(1+d_{m_i}\phi\right)m_i(\Lambda_3) \label{eq:quark_mass_variation}\, .
\end{align}
If one of these couplings is non-zero, the corresponding parameter would not be constant (in space and/or time) and will become depndent on the value of the field $\phi$. This property is interpretated as a violation of the local position invariance, or more generally of the Einstein equivalence principle.  

Following the work of \cite{damour:2010zr}, we introduce the mean light quark mass $\hat{m} = (m_u+m_d)/2$ and the difference of light quark masses $\delta m = m_d-m_u$. Following Eq.~\eqref{eq:quark_mass_variation}, the numerical value of these new parameters will change as
\begin{align}
\hat{m}(\phi) &= \left(1+d_{\hat{m}}\phi\right)\hat{m} \, , \\
\delta m(\phi) &= \left(1+d_{\delta m}\phi\right)\delta m \, ,
\end{align}
with 
\begin{align}
d_{\hat{m}} &= \frac{m_u d_{m_u}+m_d d_{m_d}}{m_u+m_d},\,\\
d_{\delta m} &= \frac{m_d d_{m_d}-m_u d_{m_u}}{m_d-m_u}.
\end{align}
\end{subequations} 
Following \cite{damour:2010zr}, we can define the dimensionless dilatonic mass charge for the body $B$ given by
\begin{align}
Q^B_d &= \frac{\partial \ln m_B(\phi)}{\partial \phi} \,  \label{dilatonic_mass_charge} ,
\end{align}
which encodes the spacetime variation of the mass of B. 
The dilatonic mass charge Eq.~\eqref{dilatonic_mass_charge} is a linear combination of the various $d_i$ couplings, i.e \cite{damour:2010zr}
\begin{subequations}
\begin{align}\label{partial_dil_mass_charge}
    Q^\mathrm{B}_d &= d_g+Q^{B}_{m_e} (d_{m_e}-d_g)+ Q^{B}_{e} d_e +\,\nonumber \\
    &Q^{B}_{\hat m}(d_{\hat m}-d_g)+Q^{B}_{\delta m}(d_{\delta m}-d_g)\, ,
\end{align}
with \cite{damour:2010zr} 
\begin{align}\label{dilatonic_partial_mass_charge}
Q^B_{\hat{m}} &= 0.093 - \frac{0.036}{A^{1/3}} - 0.02\frac{(A-2Z)^2}{A^2} -\,\nonumber \\
&1.4 \times 10^{-4} \frac{Z(Z-1)}{A^{4/3}} \, , \\
Q^B_{\delta m} &= 0.0017\frac{A-2Z}{A} \, , \\
Q^B_{m_e} &= 5.5\times 10^{-4} \frac{Z}{A}\, , \\
Q^B_{e} &= \left(-1.4 + 8.2 \frac{Z}{A} + 7.7\frac{Z(Z-1)}{A^{4/3}}\right)\times 10^{-4} \, ,
\end{align}
\end{subequations}
where $A$ and $Z$ are respectively the mass and charge numbers of $B$.

At cosmological scale, the dilaton behaves as a classical oscillating field
\begin{subequations}
\begin{align}\label{eq:dilaton_field}
\phi = \phi_0 \cos(\omega_\phi t - \vec k_\phi \cdot \vec x + \Phi) \, ,    
\end{align}
where 
\begin{align}\label{eq:amp_scalar_field}
    \phi_0 &= \frac{\sqrt{8 \pi G \rho_\mathrm{DM}}}{\omega_\phi c} \, ,
\end{align}
with $\rho_\mathrm{DM} = 0.4$ GeV/cm$^3$ \cite{Cirelli24}, if the dilaton field is identified as DM.
In the galactic frame, which is assumed to be the local DM rest frame, the gradient of the field in Eq.~\eqref{eq:dilaton_field} disappears. However, we will work in the heliocentric frame, which travels through the galactic halo with a velocity $\sim 10^{-3} \: c$ \cite{Evans19,Cirelli24}, such that 
\begin{align}\label{eq:k_phi}
    \vec k_\phi c^2 &= -\omega_\phi \vec v_\mathrm{DM} \, ,
\end{align}
where $v_\mathrm{DM}$ is a stochastic variable whose mean is $\sim 10^{-3} \: c$ and with a similar dispersion.
\end{subequations}

In this frame, the rest mass of the body $B$ oscillates as 
\begin{subequations}
\begin{align}\label{eq:mass_osc_dilaton}
    m_B(t,\vec x)&=m^0_B\left(1+\frac{\delta m^\phi_B}{m^0_B}\cos(\omega_\phi t - \vec k_\phi \cdot \vec x +\Phi)\right) \, , 
\end{align}
where $m^0_B$ is the unperturbed rest mass of $B$ and 
\begin{align}
    \delta m^\phi_B = \frac{\sqrt{8 \pi G \rho_\mathrm{DM}}m^0_BQ^B_d}{\omega_\phi c} \, .
\end{align}
\end{subequations}

In flat spacetime, the macroscopic Lagrangian describing the motion of an ensemble of particles with rest mass energy $m_Bc^2$ is \cite{damour:2010zr, hees18} 
\begin{equation}
    \mathcal{L}_B = -m_B(t, \vec x)c^2\left(1-\frac{v_B^2}{2c^2}\right) \, ,
    \label{macro_lagrangian}
\end{equation}
to first order in $(v_B/c)^2$ (where $v_B$ is the non-relativistic coordinate velocity of the body $B$).
Using Eq.~\eqref{eq:mass_osc_dilaton}, a simple Euler-Lagrange derivation of Eq.~\eqref{macro_lagrangian} leads, to first order, to
\begin{align}
    \vec a_B &= \left[\omega_\phi \vec v_B-\vec k_\phi c^2\right]\frac{\sqrt{8 \pi G \rho_\mathrm{DM}}}{\omega_\phi c}Q^B_d\times \,\nonumber\\
    &\sin(\omega_\phi t - \vec k_\phi \cdot \vec x +\Phi) .
    \label{EP_viol_acc_dil}
\end{align}

\subsection{Axion-Like Particles}\label{sec:ALP}

We now consider the coupling between the dimensionless pseudo-scalar ALP $a$ and gluons with coupling $1/f_a$ (in GeV$^{-1}$) which reads \cite{Kim:2022ype}
\begin{equation}\label{eq:int_axion_gluon}
	\mathcal L_\mathrm{int} = E_P\frac{g^2_3}{32 \pi^2} \frac{a}{f_a} G^a_{\mu\nu} \tilde G^{a,\mu\nu} \equiv \frac{g^2_3}{32 \pi^2} \theta G^a_{\mu\nu} \tilde G^{a,\mu\nu}\, ,
\end{equation}
where $E_P = \sqrt{\hbar c^5/8\pi G}$ is the reduced Planck energy and where $\theta = a/f_a$.
In \cite{Kim:2022ype}, it is shown that the interaction Lagrangian from Eq.~\eqref{eq:int_axion_gluon} induces a dependency of the mass of pions to the axion field, which implies a dependency of the mass of nucleons and atomic binding energy on the axion field. 

As a consequence, the rest mass of any atom will also depend on the axion field and its coupling strength with gluons. In this sense, we can define the dimensionless axionic mass charge of a body B as \cite{Gue:2024onx}
\begin{subequations}
\begin{align}
    Q^\mathrm{B}_a &= \frac{\partial \ln m_\mathrm{B}}{\partial \left(\theta^2\right)} \,  ,
    \label{axionic_charge}
\end{align}
where \cite{Gue:2024onx}
\begin{align}\label{axionic_mass_charge}
Q^\mathrm{B}_a &\approx -0.070+\left(\frac{3.98}{A^{1/3}}+2.22\frac{(A-2Z)^2}{A^2}\right.\,\nonumber \\
&\left.+1.50\frac{Z(Z-1)}{A^{4/3}}\times 10^{-2}\right)\times 10^{-3}\, .
\end{align}
\end{subequations}
If the axion field is identified as cosmological DM, i.e. if it behaves as Eq.~\eqref{eq:dilaton_field}\footnote{Note that recent works \cite{Banerjee25,DelCastillo25} showed that the coupling with QCD through Eq.~\eqref{eq:int_axion_gluon} induces a change of the axion field amplitude and gradient, as the field is also impacted by massive bodies. In the case of LISA, the perturbation from the Sun is the most relevant and an  analysis similar to then one from \cite{Gue25} would be needed to capture this effect. However, this will change only by a factor $\mathcal{O}(1)$ the estimates made on this paper, so we neglect it here.},  the rest mass of any atom will oscillate, similarly as in the dilaton case
\begin{subequations}
\begin{align}
m_B(t,\vec x)&=m^0_B\left(1+\frac{\delta m^a_B}{m^0_B}\cos(2\omega_a t-2\vec k_a \cdot \vec x +2\Phi))\right) \, ,
\end{align}
where 
\begin{align}
    \delta m^a_B &= \frac{8 \pi G \rho_\mathrm{DM} E^2_P Q^B_a m^0_B}{f^2_a \omega^2_a c^2} \, .
\end{align}
\end{subequations}
The perturbation oscillates at twice the axion field frequency because the mass depends quadratically on the axion, see Eq.~(\ref{axionic_charge}).

Similarly to the previous section, this oscillation of the mass induces an acceleration of the form
\begin{align}\label{EP_viol_acc_axion}
    \vec a_B &= \left[\omega_a \vec v_B-\vec k_a c^2\right]\frac{16 \pi G \rho_\mathrm{DM} v_\mathrm{DM} E^2_P Q^B_a}{f^2_a \omega^2_a c^2}\,\nonumber \\
    &\sin(2\omega_a t -2\vec k_a \cdot \vec x+2\Phi) .
\end{align}

\section{Theoretical signatures in space-based GW detectors}\label{sec:LISA_signatures}
We have shown in the previous section that ULDM can induce an oscillating acceleration on free-falling test masses. This implies that all test masses of the same composition will oscillate with the same amplitude and a phase that is nearly the same (the $\vec k\cdot \vec x$ term in the phase being negligible for the parameters considered in this study). The principle of space-based interferometry is to measure the Doppler shift between two separated test-masses using laser-links. Because of the light travel time between the two test-masses, this interferometry measurement will probe the oscillatory behavior of the test masses at a different phase and generate a signal. This 1-link response is derived for both the ULDM and the GW signal in Sec.~\ref{sec:1link}. Furthermore, space based GW data analysis is performed on TDI variables which are constructed to reduce the otherwise overwhelming laser frequency noise \cite{tinto:2021aa}. In Sec.~\ref{sec:TDI}, we introduce the expression of those TDI variables as well as their transfer functions (TF).

\subsection{One-link signals and transfer functions}\label{sec:1link}

The two oscillating accelerations Eqs.~\eqref{EP_viol_acc_dil} and \eqref{EP_viol_acc_axion} are atom-dependent and thus violate UFF. Therefore, one could use tests of the UFF (classical or quantum) to try to detect such oscillations (see e.g. \cite{Gue:2024onx}). However, one could still see such oscillations with an experiment involving only one atomic species, such as space-based GW detectors, due to the finite light travel time. 
In this section, we derive the signatures induced by such oscillations in LISA, and we compare with monochromatic GW signatures.

\subsubsection{Scalar ultralight dark matter}\label{sec:ULDM_1link}

We will derive in details the 1-link response function for the interactions between SM and a pure dilatonic field $\phi$ (see Sec.~\ref{sec:dilaton_lin}) and subsequently generalize it  to the ALP as well. 

The acceleration of the test mass $B$ in a generic frame is given by Eq.~\eqref{EP_viol_acc_dil}.
In the case of LISA, in the barycentric frame, the magnitude of velocity of the test mass is roughly the same as the Earth velocity around the Sun, i.e. $v_B \sim 3 \times 10^4$ m/s, which is one order of magnitude smaller than the galactic velocity. For this reason, the second term of Eq.~\eqref{EP_viol_acc_dil} $\propto -\vec k_\phi c^2 = \omega_\phi \vec v_\mathrm{DM}$ dominates and one can neglect the contribution from the first term such that the acceleration from Eq.~\eqref{eq:k_phi} simplifies to
\begin{align}
    \vec a_B &= \frac{\sqrt{8 \pi G \rho_\mathrm{DM}}\vec v_\mathrm{DM}}{c} Q^B_d \sin(\omega_\phi t - \vec k_\phi \cdot \vec x +\Phi) \, \label{EP_viol_acc_rm} .
\end{align}
Integrating once Eq.~\eqref{EP_viol_acc_rm}, we find the perturbed velocity of the test mass $B$ to be (see Appendix.~\ref{ap:delta_xA} for details)
\begin{align}\label{osc_pos_LISA}
    \delta \vec v_B(t,\vec x) &\approx \frac{-\sqrt{8 \pi G \rho_\mathrm{DM}}\vec v_\mathrm{DM}}{\omega_\phi c} Q^B_d \cos(\omega_\phi t - \vec k_\phi \cdot \vec x +\Phi) \, .
\end{align}
We will work with the relative frequency fluctuations of single-link from emitter $e$ to receiver $r$, i.e Doppler shifts, defined as
\begin{align}
    y_{re}(t) &= -\frac{1}{c}\frac{d\delta L_{re}}{dt} \, ,
\end{align}
where $\delta L_{re}$ is the length variation of the link.
Assuming that spacecraft $e$ (at position $\vec x_e$ at the emission time) sends a signal to spacecraft $r$ (at position $\vec x_r$ at reception time), the ULDM contribution to the one-way Doppler shift of light associated to the oscillation of the two test masses is obtained by projecting  the DM induced velocity difference of the two test masses $\delta \vec v$ onto the corresponding LISA arm  i.e. \cite{Yu23}
\begin{subequations}
\begin{align}
    y_{re}(t) &= -\frac{1}{c}\left(\hat n_{re} \cdot \left(\delta \vec v_r(t,\vec x_r)-\delta \vec v_e(t-L/c,\vec x_e)\right)\right) \, ,
\end{align}
where $\hat n_{re}$ is a unit vector pointing from $e$ to $r$, and where $L$ is the distance between the spacecrafts (which means we are using the constant equal armlength approximation). 

We can now express the Doppler shift induced by the dilaton-SM interaction from Eq.~\eqref{osc_pos_LISA} 
\begin{align}
    y^d_\mathrm{re}(t) &= \left(\hat n_\mathrm{re} \cdot \hat e_v\right)\frac{\sqrt{8 \pi G \rho_\mathrm{DM}}v_\mathrm{DM}Q_d}{\omega_\phi c^2}\times \, \label{eq:Doppler_scalar_DM} \\
    &\Re\left(e^{i\left(\omega_\phi t - \vec k_\phi \cdot \vec x_r + \Phi\right)}- e^{i\left(\omega_\phi t - \omega_\phi L/c-\vec k_\phi \cdot \vec x_e + \Phi\right)}\right) \, \nonumber .
\end{align}
A similar calculation in the case of ALP with  axion-gluon interaction (see Sec.~\ref{sec:ALP}) which induces an acceleration given by Eq.~\eqref{EP_viol_acc_axion} leads to a Doppler shifts which reads
\begin{align}
    y^a_\mathrm{re}(t) &= \left(\hat n_\mathrm{re} \cdot \hat e_v\right)\frac{8 \pi G \rho_\mathrm{DM} v_\mathrm{DM} E^2_PQ_a}{f^2_a \omega^2_a c^3}\times \,\label{eq:Doppler_pseudoscalar_DM} \\
    &\Re\left(e^{2i\left(\omega_a t -\vec k_a \cdot \vec x_r + \Phi\right)}-  e^{2i\left(\omega_a t - \omega_a L/c - \vec k_a \cdot \vec x_e+ \Phi\right)}\right) \, \nonumber .
\end{align}
\end{subequations}
In these expressions,  all the test masses are of the same composition and therefore have the same mass charge $[Q_M]_d$ (we removed the unnecessary superscript) and we defined $\vec v_\mathrm{DM} = v_\mathrm{DM}\hat e_v$. 

Following \cite{Yu23}, the transfer function (TF) of the one-arm Doppler shift induced by the pure DM scalar field is defined as the Fourier transform of $y_{re}(t)$ Eq.~\eqref{eq:Doppler_scalar_DM} normalized by the constant amplitude
\begin{align}\label{eq:TF_DM}
    \mathcal{T}^\mathrm{DM}_\mathrm{re}(\omega) &=\left(\hat n_{re}\cdot \hat e_v\right)\delta(\omega-\omega_\phi)\, \nonumber \\
    &\Re\left(e^{-i\vec k_\phi \cdot \vec x_r }-e^{-i\left(\omega_\phi L/c+\vec k_\phi\cdot \vec x_e\right)}\right)\, ,
\end{align}
where $\vec k_\phi \equiv \omega |\vec v_\mathrm{DM}|\hat k_\phi/c^2$. 
\subsubsection{Monochromatic gravitational waves}\label{sec:GW_1link}
We model the GW with frequency $f_\mathrm{GW}= \omega_\mathrm{GW}/2\pi$ in the source frame propagating along the $\hat k$ direction as \cite{LDC_tech_note}
\begin{subequations}
\begin{align}
    h_{\mu\nu}(t,\vec x) = \Re\left[\left(A_+\epsilon^+_{\mu\nu} -i A_\times \epsilon^\times_{\mu\nu}\right)e^{i\varphi(\xi)}\right] \, ,
\end{align}
where $\xi = t - \hat k \cdot \vec x/c$ represents the surfaces of constant phase, $\epsilon^+_{\mu\nu},\epsilon^\times_{\mu\nu}$ are the polarization tensors in the source frame, $A_{+,\times}$ are the amplitudes of $
+$ and $\times$ polarizations, and   
\begin{align}
    \varphi(\xi) &= \omega_\mathrm{GW} \xi + \frac{1}{2}\dot \omega_\mathrm{GW}\xi^2+\Phi_\mathrm{GW} \, ,
\end{align}
\end{subequations}
is the phase of the wave, where $\Phi_\mathrm{GW}$ is a constant and where we consider a possible frequency drift $\dot \omega_\mathrm{GW}$. 

In the barycentric reference frame, we define the triad $(\hat u, \hat v, \hat k)$, depending on the source location parametrized by its ecliptic latitude $\beta$ and ecliptic longitude $\lambda$, as \cite{LDC_tech_note,Petiteau08}
\begin{align}
    \hat u &= \begin{pmatrix}
         -\sin(\beta)\cos(\lambda)\\
        -\sin(\beta)\sin(\lambda)\\
        \cos(\beta)
    \end{pmatrix}\: , \:
    \hat v = \begin{pmatrix}
        \sin(\lambda)\\
        -\cos(\lambda)\\
        0
    \end{pmatrix}\: , \: 
    \hat k = -\hat u \times \hat v \, .
\end{align}
The polarization tensors in the barycentric frame are rotated by the polarization angle $\Psi$ i.e \cite{Cornish03}
\begin{subequations}
\begin{align}
e^+_{\mu\nu} &= \hat u \otimes \hat u-\hat v \otimes \hat v = \cos(2\psi)\epsilon^+_{\mu\nu}+\sin(2\psi)\epsilon^\times_{\mu\nu} \, \\
e^\times_{\mu\nu} &= \hat u \otimes \hat v+\hat v \otimes \hat u = -\sin(2\psi)\epsilon^+_{\mu\nu}+\cos(2\psi)\epsilon^\times_{\mu\nu} \, .
\end{align}
\end{subequations}
Finally, in the barycentric frame, the observed $+$ and $\times$ strain depend on the inclination of the source $\imath$ as (at leading order in post-Newtonian expansion) \cite{Blanchet96}
\begin{align}
    A_+ &= -\mathcal{A}(1+\cos^2(\imath)) \quad , \quad A_\times &= -2\mathcal{A}\cos(\imath)\, ,
\end{align}
where $\mathcal{A}$ depends only on source parameters \cite{Blanchet96}.
This allows us to express the polarization tensor in the barycentric frame as 
\begin{align}
    h_{ij}(t,\vec x) = \Re\left[\mathcal{A}\hat h^\mathrm{SSB}_{ij}e^{i\varphi(\xi)}\right] \, ,
\end{align}
where 
\begin{subequations}
\begin{align}
    \hat h^\mathrm{SSB}_{ij} &= (u_i u_j-v_iv_j)\hat h^\mathrm{SSB}_++(u_iv_j+u_jv_i)\hat h^\mathrm{SSB}_\times  \,\\
    \hat h^\mathrm{SSB}_+ &= -\cos(2\Psi)\left(1+\cos^2(\imath)\right)+2i\sin(2\Psi)\cos(\imath) \, , \\
    \hat h^\mathrm{SSB}_\times &= \sin(2\Psi)\left(1+\cos^2(\imath)\right)+2i\cos(2\Psi)\cos(\imath) \, .
\end{align} 
\end{subequations}
In the barycentric frame, the one-arm Doppler shift induced by the GW  can be written as 
\cite{Cornish03,LDC_tech_note,Petiteau08}
\begin{subequations}
\begin{align}
    &y^\mathrm{GW}_\mathrm{re} = \frac{-n^i_\mathrm{re} \hat n^j_\mathrm{re}}{2\left(1-\hat n_\mathrm{re} \cdot \hat k \right)}\left(h_{ij}(\xi_r)-h_{ij}(\xi_e)\right) \, \\
    &\equiv \frac{-\mathcal{A}\hat n^i_\mathrm{re} \hat n^j_\mathrm{re}}{2\left(1-\hat n_\mathrm{re} \cdot \hat k \right)}\Re\left[\hat h^\mathrm{SSB}_{ij}\left(e^{i\varphi(\xi_r)}-e^{i\varphi(\xi_e)}\right)\right] \label{eq:Doppler_GW} \, ,
\end{align}
\end{subequations}
where $\xi_r,\xi_e$ are respectively evaluated at the time and position of reception, and time and position of emission. 

Then, in the constant equal armlength approximation, the one-link transfer function is (assuming $\dot \omega_\mathrm{GW}=0$ for simplicity)
 \begin{align}\label{eq:TF_GW}
    \mathcal{T}^\mathrm{GW}_\mathrm{re}(\omega)&=\frac{-\hat n^i_\mathrm{re} \hat n^j_\mathrm{re}}{\left(1-\hat n_\mathrm{re} \cdot \hat k \right)}\delta(\omega-\omega_\mathrm{GW})\Re\left[\hat h^\mathrm{SSB}_{ij}\times \right. \, \nonumber \\
    &\left.\left(e^{-i\vec k_\mathrm{GW}\cdot \vec x_r}-e^{-i(\omega_\mathrm{GW}L/c+\vec k_\mathrm{GW}\cdot \vec x_e)}\right)\right] \, ,
\end{align}
where $\vec k_\mathrm{GW}= \omega_\mathrm{GW}\hat k_\mathrm{GW}/c$.

We have derived the one-link Doppler shift induced by scalar DM on one hand and GW on the other hand. We will extend these results to full interferometric combinations used in LISA in the next section.

\subsection{Time Delay Interferometry}\label{sec:TDI}

In space-based GW detectors, the laser noise on one-link measurement is expected to dominate largely the signal by at least 8 orders of magnitude, which would make the GW detection impossible. Fortunately, there exists a method to effectively lower drastically the laser noise by combining signals from the different S/C in such a way that the resulting observable is independent of laser noise, but still contains the GW signal (or the DM signal in our case). This is called \textit{Time Delay Interferometry}, or TDI and it was first introduced in \cite{Armstrong99, Tinto99, Dhurandhar02}. The principle relies on the fact that each laser noise is measured several times. Some specific time retarded signals combinations between the six different interferometers strongly reduce the laser noise, whilst keeping the GW (or DM) signal, thus the name of the method.

We first present the various TDI combinations that will be used for the analysis in the following sections and that are the TDI combinations currently envisioned in standard LISA pipeline. We will work exclusively with second generation TDI.
The first generation combinations are not sensitive to the laser noise in the constant armlength approximation while the second generation takes into account effects of rotation of the detector (Sagnac effect) and of the time variation of armlengths. The residuals of laser noise in the second generation TDI is proportional to the difference in spacecraft velocities squared and to terms related to S/C acceleration \cite{Hartwig22}.

Following \cite{Bayle21,Nam23},  the second generation Michelson $X_2$ combination writes as
\begin{align}\label{eq:TDI_X_gen}
    &X_2 = (1-D_{12131})\left(y_{13}+D_{13}y_{31}+D_{131}y_{12}+D_{1312}y_{21}\right)\,\nonumber\\
    &-(1-D_{13121})\left(y_{12}+D_{12}y_{21}+D_{121}y_{13}+D_{1213}y_{31}\right) \, ,
\end{align}
with subscripts denoting satellites and where we used a compacted notation for the delay operators $D_{ij}y_{jk}(t) = y_{jk}(t-L_{ij}/c)$ and $D_{ijk}y_{kl}(t) = y_{kl}(t-L_{ij}/c-L_{jk}(t-L_{ij}/c))$.  The other two Michelson $Y_2$ and $Z_2$ combinations can be obtained from Eq.~\eqref{eq:TDI_X_gen} by a cyclic permutation of the indices $1 \rightarrow 2 \rightarrow 3 \rightarrow 1$.  It can be shown geometrically that the circulation of the virtual laser beams in that case resembles an usual optical Michelson interferometers, thus the name of the combination. 

In the constant armlength approximation, Fourier transform (denoted with a hat) of the second generation TDI X combination Eq.~\eqref{eq:TDI_X_gen} can be written as
\begin{align}
    \hat X_2(\omega) &= -4\sin \left(\frac{\omega L}{c}\right)\sin \left(\frac{2\omega L}{c}\right) e^{-3i\omega L/c} \times \,\label{eq:TDI_X2_approx} \\
    &\left(\hat y_{13}(\omega) -\hat y_{12}(\omega) + e^{-i\omega L/c}\left(\hat y_{31}(\omega)-\hat y_{21}(\omega)\right)\right)\, \nonumber ,
\end{align}
where we used the fact that one delay operator corresponds to a multiplicative factor $\exp(-i\omega L/c)$ in the Fourier domain. This expression is general and can be used for both DM and GW signals.

Using the single links transfer functions derived in the previous section, we now derive the transfer function of the  full TDI combinations. 
In Appendix.~\ref{ap:TF}, we compute explicitly the transfer function of the second generation $X_2$ TDI combination in the constant armlength approximation for scalar DM and for a monochromatic GW (i.e. assuming $\dot f = 0$). For the former, one has
\begin{subequations}
\begin{align}\label{eq:amp_TF_X2}
    &\left|\mathcal{T}^\mathrm{DM}_X(\omega)\right| = 16\left|\hat n_{23}\cdot \hat e_v\sin \left(\frac{\omega L}{c}\right)\sin \left(\frac{2\omega L}{c}\right)\right|\sin^2\left(\frac{\omega L}{2c}\right) \, .
\end{align}
In the long wavelength approximation, i.e. when $\omega L/c \ll 2\pi$, i.e $f \leq 0.1$ Hz, the amplitude of the transfer function reads
\begin{align}\label{eq:TDI_X2_DM_low_freq}
    \left|\mathcal{T}^\mathrm{DM}_X(\omega)\right| &\approx 8\left(\frac{\omega L}{c}\right)^4 \left|\hat n_{23}\cdot \hat e_v\right| \, ,
\end{align}
\end{subequations}
i.e the transfer function scales as $f^{4}$. We do not recover the scaling of the transfer function of \cite{Yu23} which scales as $f^{3}$, because Eq.~\eqref{eq:amp_TF_X2} corresponds to the amplitude of the transfer function of the second generation TDI, while \cite{Yu23} used the 1.5 generation for their calculation. However, one can show easily that the transfer function of the first generation TDI is $\left|\mathcal{T}^\mathrm{DM}_X(\omega)\right|/2\sin (2\omega L/c)$, which allows us to recover the results of \cite{Yu23}.

\begin{figure}
    \centering
    \includegraphics[width=0.5\textwidth]{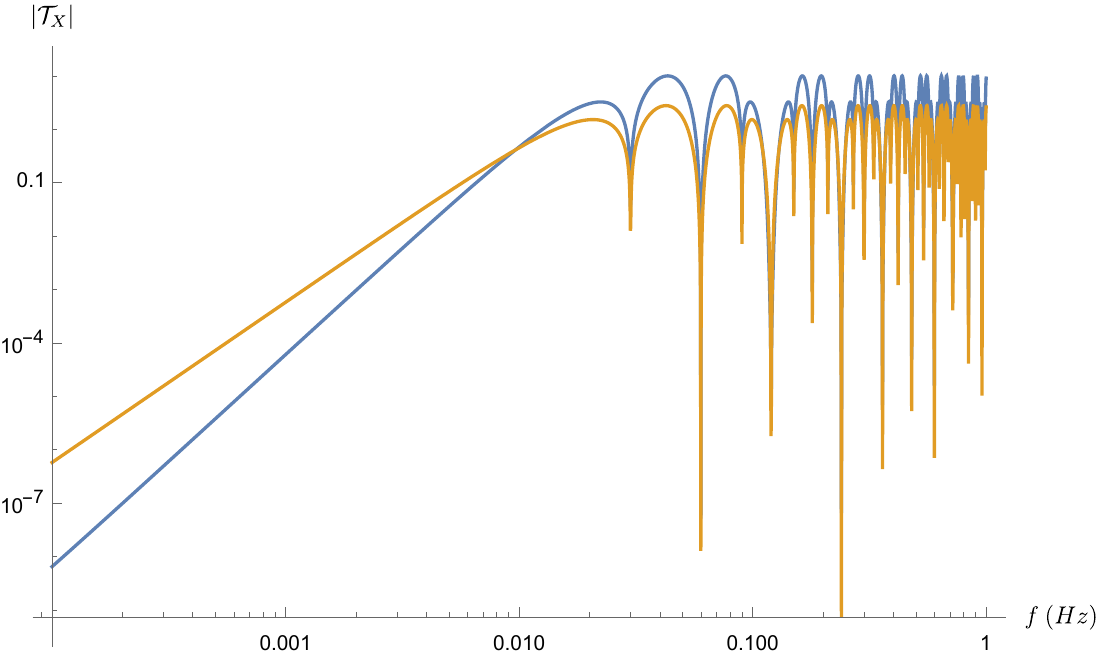}
    \caption{Amplitude of transfer functions of TDI $X$ combinations for both scalar ULDM (in blue) and GW (in orange) from Eqs.~\eqref{eq:amp_TF_X2} and \eqref{eq:TDI2_GW_TF} respectively. For this plot, we neglected the impact of the geometric factors.}
    \label{fig:TF_LISA}
\end{figure}

For the GW, the complete form of the TF can be found in App.~\ref{ap:TF}. When $\omega L/c \ll 2\pi$, one can obtain a concise expression of such TF which reads (see App.~\ref{ap:TF} for a detailed derivation)
\begin{align}\label{eq:TDI_X2_GW_low_freq}
    |\mathcal{T}^\mathrm{GW}_X(\omega)| &= 8\left(\frac{\omega L}{c}\right)^3\left|\hat h^\mathrm{SSB}_{ij} \left(\hat n^i_{13}\hat n^j_{13}-\hat n^i_{12}\hat n^j_{12}\right)\right| \, .
\end{align}
A comparison of Eqs.~\eqref{eq:TDI_X2_DM_low_freq} and \eqref{eq:TDI_X2_GW_low_freq} shows that  the DM signal will, on average (i.e neglecting the geometric factor), be suppressed by a factor $\omega L/c$ compared to the GW signal at low frequencies, see Fig.~\ref{fig:TF_LISA}.
This is due to the fact that the DM signal is a dipolar signal which travels at a non-relativistic velocity while the GW signal is a quadrupolar signal propagating at the speed of light. The full derivation of the DM TF provided in App.~\ref{ap:TF} shows that it contains another term which behaves as $\sim (\omega L/c)^3 v_\mathrm{DM}/c$ at low frequency which becomes dominant for large $v_\mathrm{DM}$. Therefore, the TF for relativistic DM would behave similarly to the GW TF at low frequencies. On the other hand, a similar calculation for a GW propagating at a non-relativistic velocity shows that the TF  also scales as $(\omega L/c)^3$ at low frequency, contrarily to the DM dipolar signal. 

This behaviour can be understood from a simple argument.  Assuming a low frequency (i.e $\omega \ll c/L \equiv 1/\tau$, where $\tau$ is the light travel time between spacecrafts which implies that the wavelength is much larger than the armlength $\lambda \gg L$) and non-relativistic wave  (velocity $v \ll c$), the phase of the wave at the location of the emitting and receiving spacecrafts is very close (i.e $\Delta \varphi \ll 2\pi$) and therefore, the one-way Doppler  is $\mathcal{O}(\omega L/c)$, see Eq.~\eqref{eq:TF_DM}. Because of the slow wave velocity, during a two-way round trip, the wave has not significantly propagated and the Doppler contribution of the two ways have the same order of magnitude and a different sign in the case of DM and the same sign in the case of the GW. Put in other words, the two-way transfer functions $\mathcal{T}^\mathrm{DM}_{131}=\mathcal{T}^\mathrm{DM}_{13}+\exp(-i\omega L/c)\mathcal{T}^\mathrm{DM}_{31}$ is $\mathcal{O}\left((\omega L/c)^2\right)$, while the two-way transfer function for the GW is still $\mathcal{O}(\omega L/c)$. This explains the impact of polarization on the TDI TF for non relativistic waves in the low frequency regime.

Now, let us consider the cases of relativistic waves. For the DM wave, the two-way Doppler cancellation observed in the non-relativistic case no longer occurs (i.e. $\mathcal T_{13}^\mathrm{DM} \sim -T_{31}^\mathrm{DM}$ is no longer valid). This is due to the fact that the Doppler phase has an additional $\vec k \cdot \vec x$ term that is non-negligible for relativistic waves, and which depends on the direction of propagation of the photon. This breaks the symmetry between the inward and outward Doppler at leading order such that, in the low frequency regime, the DM signal becomes $\mathcal{O}(\omega L/c)$ as well. For relativistic GW, the reasoning is similar to the ones for non-relativistic quadrupolar waves: both one-way Doppler terms are $\mathcal O(\omega L/c)$ at low frequency. Due to the quadrupolar nature of the wave, both these contributions add up in the 2-way Doppler which also has a $\mathcal O(\omega L/c)$ behavior at low frequency.

% \jg{Now, let us assume that both waves are relativistic. For the GW, still considering a round trip between two spacecrafts, the difference between the inward and outward Doppler will be of the same order as in the non-relativistic case. Indeed, in this situation, the phase of the Doppler picks up a contribution which depends on the relative orientation between the GW wave vector and the photon propagation direction. Therefore, the signal is still $\mathcal{O}(\omega L/c)$ at low frequency. For the DM, a similar change happens : the phase of the signal in one way is different from the phase in the other way. Specifically, the phase difference between the inward and the outward effect picks up an additional term (compared to the non relativistic case) $\vec k \cdot (\vec x_r - \vec x_e)=\mathcal{O}(\omega L/c)$ and crucially, this term depends on the direction of propagation of the photon. This breaks the symmetry between the inward and outward Doppler at leading order such that, in the low frequency regime, the DM signal becomes $\mathcal{O}(\omega L/c)$ as well.} 

Those arguments explain why DM, as a non relativistic wave and with a dipolar effect on LISA arm, has not the same transfer function as GW at low frequency.

Other important TDI combinations are the $A,E,T$ combinations which are linear combinations of $X,Y,Z$ \cite{Prince02}
\begin{subequations}\label{eq:AET_TDI}
    \begin{align}
        A(t) &= \frac{1}{\sqrt{2}}\left(Z(t) - X(t)\right) \, \\
        E(t) &= \frac{1}{\sqrt{6}}\left(X(t)-2Y(t)+Z(t)\right)\,\\
        T(t) &= \frac{1}{\sqrt{3}}\left(X(t)+Y(t)+Z(t)\right) \, ,
    \end{align}
\end{subequations}
which are used in data analysis. These combinations are useful because their cross noise power spectral density are vanishing in the equal armlength approximation \cite{Nam23} and they form a set of quasi orthogonal channels. 

We have now derived the signatures of scalar DM and monochromatic GW on LISA TDI combinations. In the next section, we will use them to build Bayesian models for scalar DM and monochromatic GW in order to compare the two on a given dataset.

\section{Bayesian inference to discriminate ULDM from GW}\label{sec:data_analysis}

We now turn to the main focus of this paper : the numerical analysis of TDI dataset containing either a monochromatic galactic binary (GB) or (scalar) ULDM.

For each simulated dataset, we will fit two models: (i) a scalar ULDM model and (ii) a GB model using a realistic LISA noise model. Using Bayesian model selection tools, we will assess if LISA data will be able to discriminate between the two types of signals. Evidently, this is an important question to answer. Indeed, many recent studies have claimed that LISA and other space-based GW detectors could probe unconstrained regions of ULDM parameter spaces, see e.g. \cite{Pierce18,Morisaki21,Yu23}. Those studies implicitly assume that the ULDM signal would be completely uncorrelated with the thousands of GW signals that LISA will detect, or in other words that the ULDM signal is totally uncorrelated with all the GW signals. This is what we want to verify through this study.

\subsection{Methodology}

\subsubsection{\label{sec:Generation_signal_LISA}Time domain generation of signals}
We first simulate a dataset which contains either a GB or scalar DM signal. To do so, we simulate in time domain the second generation TDI Eqs.~\eqref{eq:AET_TDI} and ~\eqref{eq:TDI_X_gen} induced by either a GB or a scalar ULDM signals through Doppler effects derived in the previous sections, see Eqs.~\eqref{eq:Doppler_scalar_DM} and \eqref{eq:Doppler_GW}. The positions of the spacecrafts are obtained using \textit{LISAOrbits} \cite{Bayle22}. We use Earth-trailing orbits \cite{Martens21},  which we simulate for one full orbit around the Sun, i.e $T_\mathrm{obs} = 365 \times 86400$ s. We generate data with a sampling frequency $f_s=0.1$ Hz, i.e $N=T_\mathrm{obs}f_s$ measurements are performed. The data are generated without noise added, such that the results of the analysis, and its possible bias, will not depend on noise.

Note that a GB signal depends on 8 parameters: amplitude ($\mathcal A$), frequency ($f$) and frequency derivative ($\dot f$), sky location ($\beta$, $\lambda$), polarisation $\Psi$, inclination $\iota$ and initial phase $\Phi$, see Sec.~\ref{sec:GW_1link}. The ULDM signal, on the other hand, depends on 6 parameters: an amplitude (which is related to the scalar-matter coupling), a frequency, the DM velocity ($v_\mathrm{DM}$), the sky location (direction the DM velocity, $\hat e_v$), and an initial phase, see \ref{sec:ULDM_1link}.

\begin{figure}
    \centering
    \includegraphics[width=0.5\textwidth]{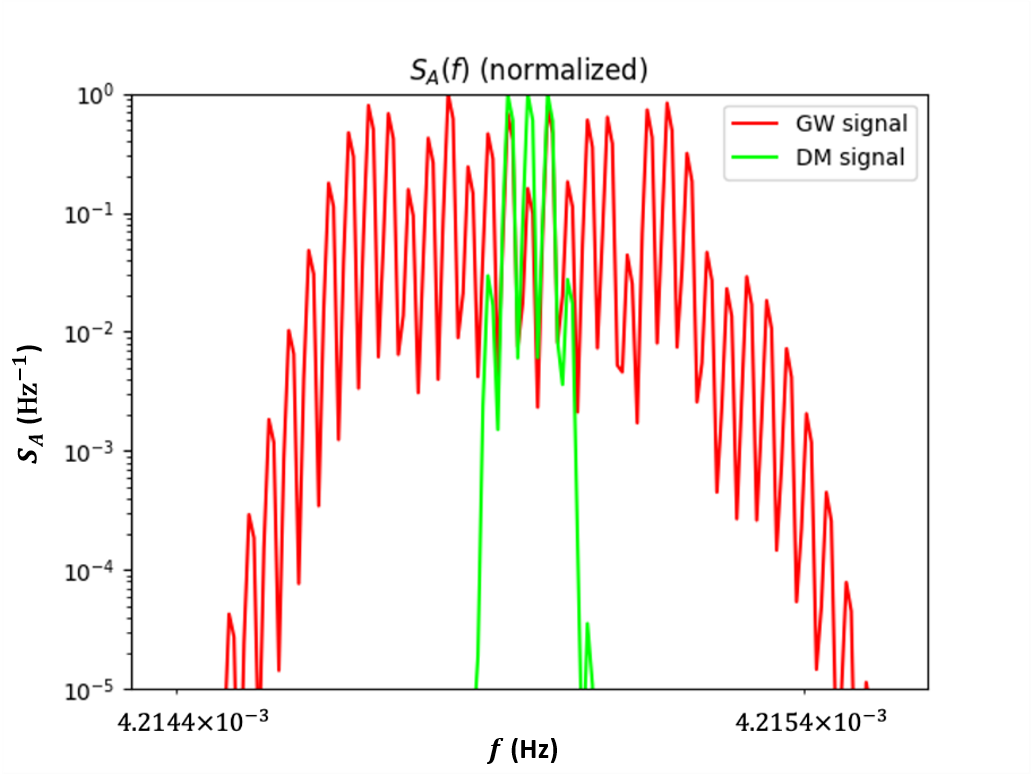}
    \caption{Normalized PSD of the DM and GW simulated signals around the signal frequency located at $f\sim 4.215$ mHz. The different harmonics are separated by the Earth orbital frequency around the Sun. This figure has been generated assuming 4 years of data.}
    \label{fig:PSD_sig_zoom}
\end{figure}

We show in Fig.~\ref{fig:PSD_sig_zoom} the PSD of DM and GW simulated signals at the same frequency $f\sim 4.215$ mHz with parameters shown in Tab.~\ref{tab:GW_GW} and \ref{tab:DM_DM}, both normalized to their maximum values. As pointed out in \cite{Yao25}, the spectral shape of DM and GW signals is different around the signal frequency due to the orbital motion of the detector.

\subsubsection{Likelihood, signal and noise modeling}
We use a Gaussian likelihood. Since our signals cover a very narrow frequency band, we express it in the frequency domain. This likelihood is given by
\begin{align}\label{eq:likelihood_general_LISA}
    - \log \mathcal{L} &= \sum_{x=A,E} \sum_{j=0}^{N/2} \Bigg[\log\left(2\pi  f_s N_{x}(f_j)\right) + \frac{|\tilde d^x_j- \tilde m^x_j|^2}{N f_sN_x(f_j)} \Bigg] \, ,
\end{align}
where $N_{A,E}$ are the noise PSD of the TDI channels, $\tilde d_j^x$ the Fourier transform of the simulated data and $\tilde m^x_j$ the Fourier transform of the model fitted to the data (that depends on the sampled parameters).

Note that the signal-to-noise ratio (SNR) for a signal is computed with \cite{Babak21}
\begin{align}\label{eq:SNR}
    \mathrm{SNR} = \sqrt{4\Re\left(\int df \frac{\tilde d_\mathrm{GW} \tilde d^\dag_\mathrm{GW}}{N(f)}\right)} \, .
\end{align}

For the noise modeling, we consider only the two dominant sources of noise in the TDI channels: the Optical Metrology System (OMS) noise and the test-mass acceleration noise. We use the expressions of the TDI noise PSD which are obtained in the constant and equal armlength approximation in \cite{Nam23}, i.e.
\begin{align}\label{eq:TDI_A_noise_PSD}
    &N_{A,E}(f) = 32\sin^2\left(\frac{2 \pi f L}{c}\right)\sin^2\left(\frac{4 \pi f L}{c}\right)\,\nonumber \\
    &\left(2\left(3+2\cos\left(\frac{2 \pi f L}{c}\right)+\cos\left(\frac{4 \pi f L}{c}\right)\right)S_\mathrm{acc}(f)+\right.\,\nonumber \\
    &\left.\left(2+\cos\left(\frac{2 \pi f L}{c}\right)\right)S_\mathrm{oms}(f)\right) \, .
\end{align}
assuming no correlation between the noises and where single-link optical metrology noise and single test mass acceleration noise respectively read \cite{Robson19,Babak21}
\begin{subequations}\label{eq:noise_PSD_LISA}
\begin{align}
S_\mathrm{oms} &= \left(1.5 \times 10^{-11} \: \mathrm{m}\right)^2\left(1+\left(\frac{2 \: \mathrm{mHz}}{f}\right)^4\right) \: \mathrm{Hz}^{-1} \, \\
S_\mathrm{acc} &= \left(3 \times 10^{-15}\: \mathrm{m.s}^{-2}\right)^2\left(1+\left(\frac{0.4 \: \mathrm{mHz}}{f}\right)^2\right)\,\nonumber\\
&\left(1+\left(\frac{f}{8 \: \mathrm{mHz}}\right)^4\right)\: \mathrm{Hz}^{-1} \, ,
\end{align}
\end{subequations}
corresponding to the instrument performance requirements \cite{Colpi24}. 

Regarding the signal modeling, i.e. $\tilde m^x_j$, we use a fast version of the TDI variable following the work from \cite{Cornish07} which has been extended to account for varying armlengths and which has also been adapted for the case of ULDM signal. The main idea of this fast TDI waveform generator (in the Fourier domain) is to use an heterodyne decomposition of the TDI signal, i.e. to express it as a product of a fast purely monochromatic term (at a frequency corresponding to one of the Fourier bin of the data) with a slowly varying signal. The Fourier transform of the slowly varying part is computed using a Discrete Fourier Transform (DFT) using a spare sampling grid, which is significantly faster than using a DFT of the full signal. The Fourier transform of the oscillating part is trivial. These fast TDI waveform generators are described in details in App. ~\ref{ap:fastDM_fastGB}.

\subsubsection{Priors}
For the GW source parameters, we use uniform prior distribution for all of them, except for the sky location parameters that are assumed to follow an isotropic distribution.

For the DM source parameters, we also use uniform prior distribution for all of them except for the sky location which follows an isotropic distribution and the DM velocity. The prior for the norm of the DM velocity ($\pi_\mathrm{DM}$) is provided  by the angular integral of the Maxwellian velocity distribution $\mathfrak{F}(\vec v)$ of the Standard Halo Model \cite{Derevianko18, foster:2018aa}
\begin{align}\label{DM_vel_distrib}
    \pi_\mathrm{DM}(v) &= \int d\Omega \mathfrak{F}(\vec v) =\int d\Omega \frac{1}{\left(2\pi \sigma^2_{v}\right)^{3/2}} e^{-\frac{\left(\vec v - \vec v_\mathrm{DM} \right)^2}{2\sigma^2_{v}}} \, ,
\end{align}
where $\sigma_v$ is the dispersion (virial) velocity and $\vec v_\mathrm{DM}$ is the mean Earth's velocity in the galactic frame, whose amplitude is $10^{-3}c$ and whose direction can be expressed in the ecliptic frame as $\hat e_v = (\cos(\beta_\mathrm{DM})\cos(\lambda_\mathrm{DM}),\cos(\beta_\mathrm{DM})\sin(\lambda_\mathrm{DM}),\sin(\beta_\mathrm{DM}))$, where $\beta_\mathrm{DM}\sim 59.91\degree$ and $\lambda_\mathrm{DM} \sim 24.67\degree$ are respectively the ecliptic latitude and longitude\footnote{The Sun velocity in the galactic halo points towards $\alpha$ Cygni, the biggest star of the Cygnus constellation \cite{Miuchi20}, which corresponds to a right ascension $\alpha_\mathrm{DM}=310.36\degree \mathrm{E}$ and declination $\delta_\mathrm{DM}=45.28\degree \mathrm{N}$ \cite{vanLeeuwen07}, in the equatorial frame. By transforming it to the ecliptic coordinate system \cite{Leinert98}, we find $(\beta_\mathrm{DM},\lambda_\mathrm{DM})=(59.91\degree,24.67\degree)$.} \footnote{Note that our choice of priors implies that we simplify the system by considering that the DM velocity amplitude $v$ is uncorrelated with the DM wind direction $(\beta,\lambda)$, while both parameters are in reality constrained together by the same distribution $\mathfrak{F}(\vec v)$. As mentioned above, we will use isotropic priors for both ecliptic latitude and longitude of the DM wind, while the full galactic velocity distribution $\mathfrak{F}(\vec v)$ provides prior information on such direction. First, we make this choice because there is no analytical expression for the angular distributions from $\mathfrak{F}(\vec v)$. Second, by doing the numerical integration of $\mathfrak{F}(\vec v)$ over the velocity amplitude, one can show that the width of the angular distributions are much larger than the posterior distributions of these parameters that the simulation provides (see below, Section ~\ref{sec:model_DM_signal_DM}), and therefore using an isotropic prior is unlikely to impact the results.}.

\subsubsection{Sampling of the parameter space, results}
We use \textit{Nessai} \cite{nessai, Williams:2021qyt,Williams:2023ppp}, a nested sampling algorithm, to sample our posterior distribution. As an output, we also have an estimate of the Bayesian evidence. The ratio of the Bayesian evidences of two models used to fit the data, i.e. the Bayes factor, is used as a criteria for model selection. Furthermore, we also use the norm of the residuals as an indicator of the goodness of fit and more precisely the signal-to-noise ratio of the residuals, computed using Eq.~(\ref{eq:SNR}) and called Residuals-To-Noise Ratio (RNR).

\subsection{Results}

\subsubsection{Fit on data that contains a GW signal}

\begin{figure}[h!]
    \centering
    \includegraphics[width=0.45\textwidth]{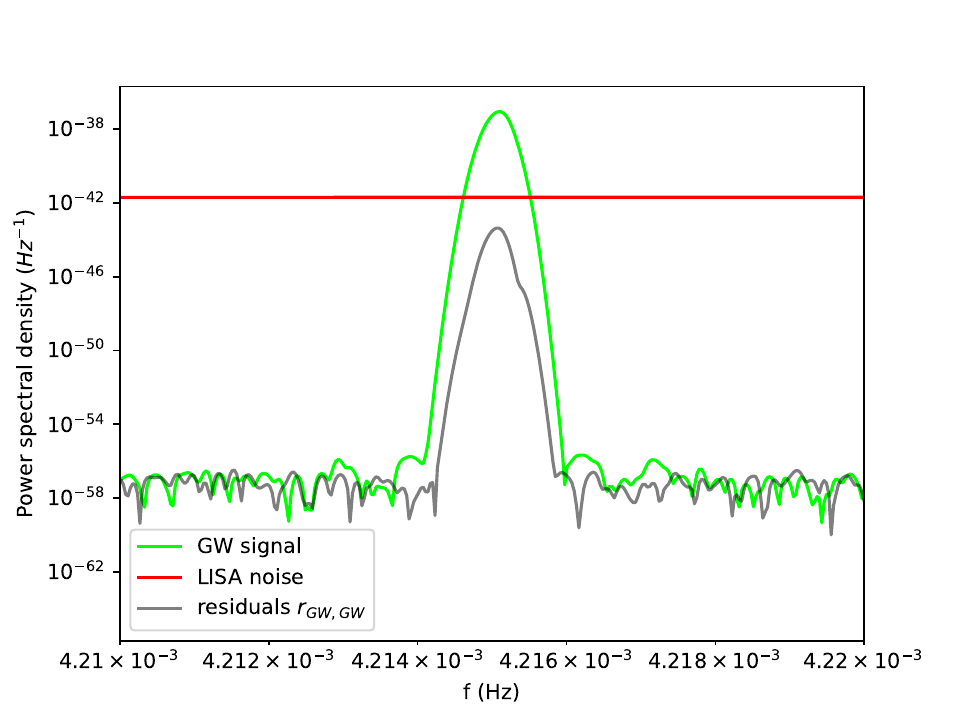}
    \caption{GW signal PSD (in lime) compared to LISA noise PSD of the TDI $A$ combination (in red) in Fourier domain. The signal is well above the noise. We also show the $r_\mathrm{GW,GW}$ residuals power in grey, which are below the noise.}
    \label{fig:GW_signal}
\end{figure}

We first generate a dataset $d_\mathrm{GW}$ that includes only a GB with parameters $D_\mathrm{GW}$ presented on the second column from Tab.~~\ref{tab:GW_GW}.  
The SNR of this dataset for both A and E TDI channel, computed using Eq.~(\ref{eq:SNR}), are 
\begin{subequations}\label{eq:SNR_GW_data}
\begin{align}
    \mathrm{SNR}^\mathrm{GW}_A &\approx 447 \, \\
    \mathrm{SNR}^\mathrm{GW}_E &\approx 480 \, ,
\end{align}
\end{subequations}
which shows that the signal is well above the noise, as  can also be noticed in Fig.~\ref{fig:GW_signal}.

\begin{table*}[t!]
    \centering
    \begin{tblr}{vlines, colspec={c||c||cc||cc||c}}
    \hline
    & GW data &\SetCell[c=2]{} GW model & &\SetCell[c=2]{} DM model & &\\
    \hline
    \hline
    GW params.& \SetCell[c=1]{} $D_\mathrm{GW}$ & \SetCell[c=1]{} $M^\mathrm{GW}_\mathrm{GW}$ & \SetCell[c=1]{} Mean $\pm$ 1 $\sigma$ &\SetCell[c=1]{} $M^\mathrm{DM}_\mathrm{GW}$ & \SetCell[c=1]{} Mean $\pm$ 1 $\sigma$ & DM params.\\
    \hline
    \hline
    $\mathcal{A}$ $(10^{-21})$ & $0.988$& $0.988$ &$ 0.988 \pm 0.003 $ & $1.406$ & $1.418 \pm 0.014$ & $Q$ $(10^{-5})$\\
    \hline
    $f$ (mHz) & $4.215$ & $4.215$ & $4.215 \pm 0.000$ & $4.215$ & $4.215 \pm 0.000$ & $f$ (mHz)\\
    \hline
    $\dot f$ (aHz/s) & $55.91$& $56.26$ &$57.36 \pm 18.20$ & $-$ & $-$ &$-$ \\
    \hline
    $\beta$ (rad) & $0.817$ & $0.8164$ & $0.8165 \pm 0.0005$ & $-0.101$ & $-0.098 \pm 0.004$ & $\beta$ (rad)\\
    \hline
    $\lambda$ (rad) & $5.149$ & $5.1485$ & $5.1485 \pm 0.0004$ & $0.806$ & $0.804 \pm 0.005$ & $\lambda$ (rad)\\
    \hline
    $\imath$ (rad) & $1.047$ & $1.048$ &$1.047 \pm 0.002$ & $-$ &$-$& $-$ \\
    \hline
    \SetCell[r=2]{}$\Psi$ (rad) & $0.785$ & $0.785$ & $0.785 \pm 0.002$ & $-$ & $-$ & $-$\\
     & & & $2.356 \pm 0.002$\\
    \hline
    \SetCell[r=2]{}$\Phi$ (rad) & $4.890$ & $4.888$ & $4.889 \pm 0.007$ & $5.521$ & $5.521 \pm 0.007$ & \SetCell[r=2]{}$\Phi$ (rad)\\
     & & & $1.748 \pm 0.007$\\
    \hline
    $-$ & $-$ & $-$ &$-$ & $1.014$ &$1.003 \pm 0.010$& $v_\mathrm{DM}$ $(10^6)$(m/s)\\
    \hline
\end{tblr}
\caption{On the first column, we show the GW injected parameters in the simulation, with the numerical values on the second column. On the third and fourth columns, we show respectively the GW model best fit, i.e the posterior with the largest likelihood, and the mean and standard deviation of the full distribution of posteriors. The GW model correctly recovers the injected parameters with less than 1\% deviation. On the right part of the table, we show the scalar DM model on the same GW data, with the corresponding scalar DM parameters. On the fifth and sixth columns, we show the model best fit and the mean and standard deviation of the full distribution of posteriors.}
\label{tab:GW_GW}
\end{table*}

\paragraph{Fit using a GW model}

\begin{figure*}
    \centering
    {\includegraphics[width=\textwidth]{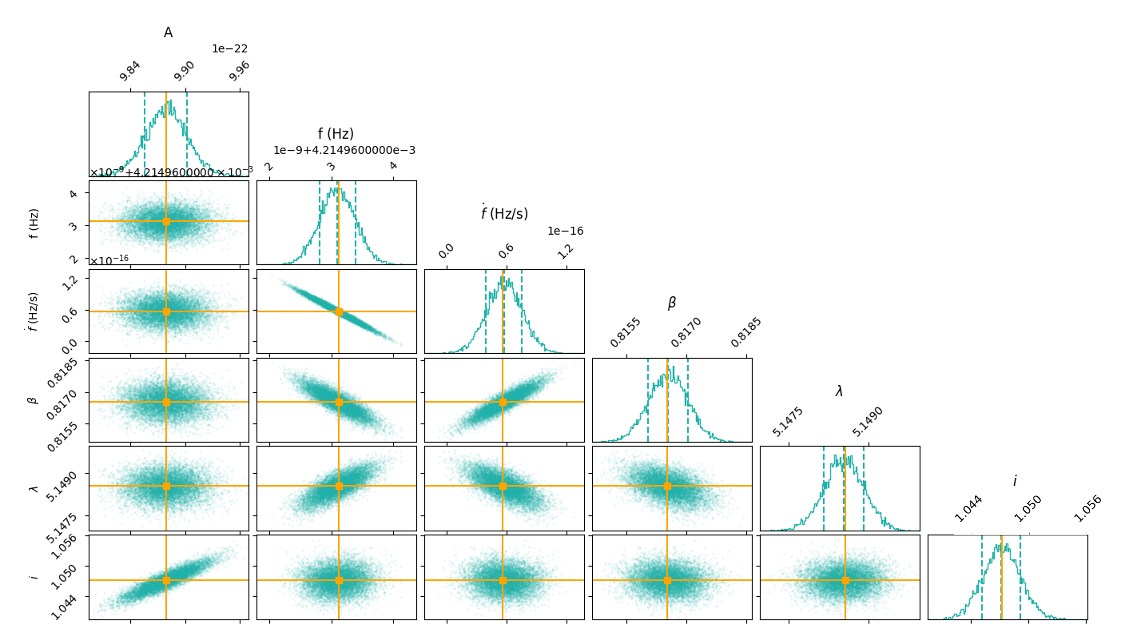}}
    \caption{2D posterior distributions of a GB model onto a GB signal with parameters from Tab.~\ref{tab:GW_GW}. The yellow lines indicate the true values of the parameters. Only data points are shown, but contours are omitted. The amplitude is in units of $10^{-22}$, the frequency is in units of $10^{-9}$ Hz with an offset of $4.21496$ mHz, and the frequency drift is shown in units of $10^{-16}$ Hz/s. For clarity, we marginalize over $\Psi$ and $\Phi$, but as can be noticed from Table \ref{tab:GW_GW}, the model best fit corresponds to the injected values (at the 1\% error level).}
    \label{fig:corner_GW_GW}
\end{figure*}
We first comment the result of the fit of a GB model onto simulated GB data. The 2D posterior distributions are shown in Fig.~\ref{fig:corner_GW_GW}. The yellow lines indicate the injected values for the parameters. One can notice that the algorithm retrieves the correct parameters.

The  best fit parameters $M^\mathrm{GW}_\mathrm{GW}$, that we define as the parameters corresponding to the largest posterior, in addition to the mean and standard deviation of the 1-D posteriors are provided in Tab.~\ref{tab:GW_GW}.

A first assessment of the goodness of the fit is done using the residuals which are defined as the difference between the simulated data and the model $m^\mathrm{GW}_\mathrm{GW}$ corresponding to the best fitted parameters $M^\mathrm{GW}_\mathrm{GW}$, i.e.
\begin{align}
        r_\mathrm{GW,GW} &= d_\mathrm{GW} - m^\mathrm{GW}_\mathrm{GW} \, .
\end{align}
We then quantify the goodness of fit using the RNR using Eq.~(\ref{eq:SNR}) on these residuals. For the A and E TDI combinations, one finds
\begin{subequations}
\begin{align}
\mathrm{RNR}^\mathrm{GW, GW}_A &\approx 0.54 \, \\
\mathrm{RNR}^\mathrm{GW, GW}_E &\approx 0.41 \, .
\end{align} 
\end{subequations}
The RNR are below unity and well below the SNR of the initial data, see Eqs.~\eqref{eq:SNR_GW_data}, indicating that the model is efficient, in the sense that the majority of the signal power is extracted from the model.
In Fig.~\ref{fig:GW_signal}, we compare the residuals to LISA noise in Fourier space. As expected from the computation of the RNR, the residuals are well below the noise.\footnote{Despite its efficiency, the residuals still present a peak at the GW frequency. This is due to the approximations done in the fast TDI waveform model (e.g. neglecting the spacecraft velocity during light travel time) presented in Appendix.~\ref{ap:fastDM_fastGB}.}.

\paragraph{Fit using a scalar DM model}

\begin{figure}[h!]
    \centering
    \includegraphics[width=0.45\textwidth]{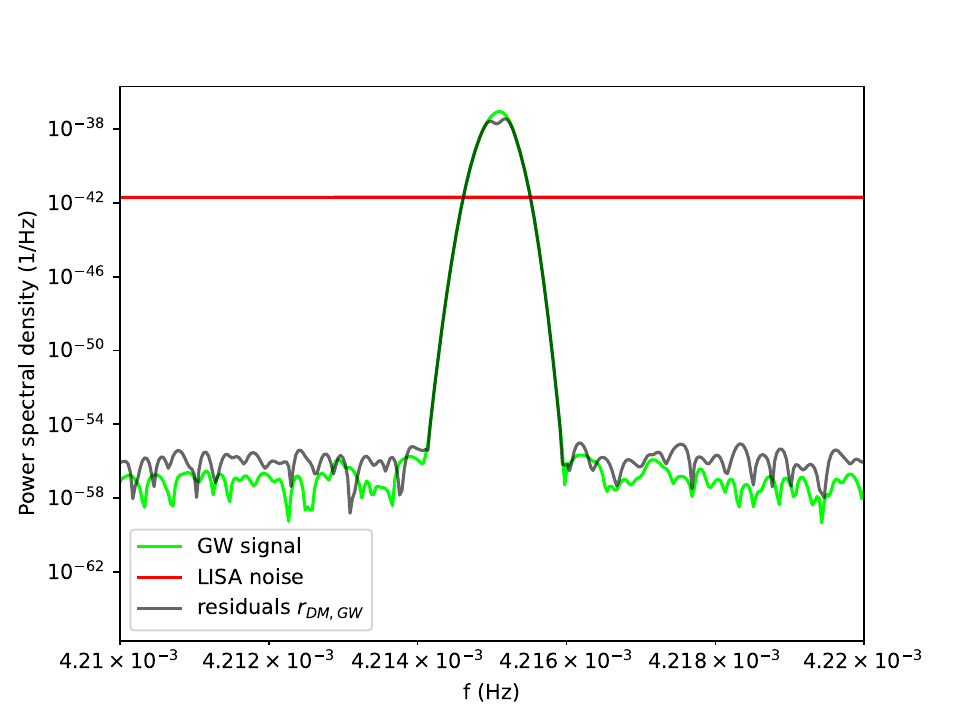}
    \caption{GW signal PSD (in lime) compared to LISA noise PSD of TDI $A$ combination (in red) in Fourier domain (as in Fig.~\ref{fig:GW_signal}). This time, the $r_\mathrm{DM,GW}$ residuals power in grey are close to the signal power and above the noise. This indicates the model fits poorly the data.}
    \label{fig:DM_GW_res_signal}
\end{figure}

We now focus on a fit using a scalar DM model on the same dataset that contains a GW signal. For this case, the resulting best fit parameters, mean and standard deviations of the posteriors are shown on the right part of Tab.~\ref{tab:GW_GW}. Here, the $v_\mathrm{DM}$ parameter denotes the norm of the velocity of the wave. In the case of a GW signal, it corresponds to the speed of light. The obtained posterior is constrained by the DM velocity prior distribution, which explains why it is driven towards $\sim 10^{-3}\: c$. 
As in the previous section, we can first compute the RNR of the best fit DM model $m^\mathrm{DM}_\mathrm{GW}$ on the GW signal and, one finds 
\begin{subequations}
\begin{align}
    \mathrm{RNR}^\mathrm{DM, GW}_A &\approx 572 \, \\
    \mathrm{RNR}^\mathrm{DM, GW}_E &\approx 682 \, .
\end{align}
\end{subequations}
The RNR values are much larger than 1, and of the same order of magnitude as the GW data SNR, see Eq.~\eqref{eq:SNR_GW_data}, implying that the model is inefficient in explaining the signal. In Fig.~\ref{fig:DM_GW_res_signal}, we compare those residuals with the injected GW signal in Fourier space. One can notice that both the residuals and the simulated data have the same spectrum amplitude, meaning that the model has not extracted any relevant information out of the data\footnote{Note that, as mentioned previously, we used the galactic velocity distribution as a prior, but for completeness, we can also relax this velocity constraint, i.e use an uniform prior in the range $[0,c]$. In this case, the best fit presents a much higher velocity (of order $10^7$ m/s), but the resulting RNR is of the same order of magnitude as above. Therefore, the result is not significantly impacted by the prior.}. 

\paragraph{Model comparison}
While in the previous sections, we evaluated the two models individually by means of residuals, we will now compare them, using a Bayesian model selection criterion : the Bayes factor $\mathcal{B}$, defined as 
\begin{align}
    \mathcal{B} &= \frac{p\left(d_\mathrm{GW}|M^\mathrm{GW}_\mathrm{GW}\right)}{p\left(d_\mathrm{GW}|M^\mathrm{DM}_\mathrm{GW}\right)} \, .
\end{align}
One can show that $\mathcal{B}$ is the ratio of the model evidences which are output from the sampler used and is given by
\begin{align}
    \log \mathcal{B} &= \mathcal{O}(10^5) \gg 1 \, .
\end{align}
This means that the GB model $M^\mathrm{GW}_\mathrm{GW}$ is largely preferred over the DM model  on this dataset which contains a simulated GW signal.

Therefore, from both residuals and Bayes factor, we can conclude that, a priori, LISA will indeed be able to make the difference between one GB and a scalar DM signal. This suggests that any individual monochromatic GW emitted by a GB will not be misinterpreted as a scalar DM in a space-based GW data analysis.

In the next section, we turn to the opposite case, i.e what happens with a scalar DM signal.

\subsubsection{Fit on data that contains scalar DM signal}\label{sec:fit_DM}

For consistency with the GB that was analyzed in the previous section, we simulate a DM signal with the same injected frequency (in order to compare to the same noise level) and intrinsic amplitude of oscillation on the test masses. From Eqs.~\eqref{eq:Doppler_scalar_DM} and \eqref{eq:Doppler_GW}, this means that we choose $Q_d$ such that
\begin{align}\label{eq:equiv_A_epsilon}
    Q_d &= \mathcal{A}\frac{2\pi f c^2}{\sqrt{8 \pi G \rho_\mathrm{DM}}v_\mathrm{DM}} \, ,
\end{align}
where $f, \mathcal{A}$ are given in Table ~\ref{tab:GW_GW}. From now on, we will discard the subscript $d$ i.e $Q_d \equiv Q$.

In addition, we will use the mean velocity amplitude and direction of the velocity distribution from Eq.~\eqref{DM_vel_distrib}. The mean amplitude is $v_\mathrm{DM}= 3 \times 10^{5}$ m/s, while the mean direction points towards $(\beta_\mathrm{DM},\lambda_\mathrm{DM})=(59.91\degree,24.67\degree)\sim(1.046 \, \mathrm{rad},0.431\, \mathrm{rad})$.
The numerical values of the parameters used for the scalar DM signal are presented in Table.~\ref{tab:DM_DM}.
\begin{table*}
\centering
    \begin{tblr}{vlines, colspec={c||c||cc||cc||c}}
        \hline
        & DM data &\SetCell[c=2]{} DM model & &\SetCell[c=2]{} GW model & &\\
        \hline
        \hline
        DM params. & \SetCell[c=1]{} $D_\mathrm{DM}$ & \SetCell[c=1]{} $M^\mathrm{DM}_\mathrm{DM}$ & \SetCell[c=1]{} Mean $\pm$ 1 $\sigma$ & \SetCell[c=1]{} $M^\mathrm{GW}_\mathrm{DM}$ & \SetCell[c=1]{} Mean $\pm$ 1 $\sigma$ & GW params.\\
        \hline
        \hline
        $Q$ $(10^{-5})$ & $2.396$ & $2.785$ &$2.299 \pm 1.289$ & $0.129$ & $0.128 \pm 0.004$ & $\mathcal{A}$ $(10^{-21})$\\
        \hline
        $f$ (mHz) & $4.215$ & $4.215$ & $4.215 \pm 0.000$& $4.215$ & $4.215 \pm 0.000$ & $f$ (mHz)\\
        \hline
        $\beta$ (rad) & $1.046$ & $1.048$ & $1.045 \pm 0.010$ & $-1.520$ & $-1.522 \pm 0.004$ & $\beta$ (rad)\\
        \hline
        $\lambda$ (rad) & $0.431$ & $0.435$ & $0.430 \pm 0.021$ & $5.326$ & $5.307 \pm 0.021$ & $\lambda$ (rad)\\
        \hline
        $v_\mathrm{DM}$ $(10^5)$ (m/s) & $2.998$ & $3.703$ & $3.659 \pm 1.239$ & $-$ &$-$ & $-$\\
        \hline
        \SetCell[r=2]{}$\Phi$ (rad) & \SetCell[r=2]{}$3.444$ & \SetCell[r=2]{}$3.444$ & \SetCell[r=2]{}$3.444 \pm 0.010$ & \SetCell[r=2]{} $4.740$ & $4.740 \pm 0.066$ & \SetCell[r=2]{} $\Phi$ (rad) \\
        & & & & & $1.597\pm 0.069$ & \\
        \hline
        $-$ & $-$ & $-$ & $-$ & $-8.040$ & $38.14 \pm 208.6$ & $\dot f$ (aHz/s) \\
        \hline
        $-$ & $-$ & $-$ & $-$ & $1.990$ & $1.995 \pm 0.024$ & $\imath$ (rad) \\
        \hline
        \SetCell[r=3]{} $-$ & $-$ & $-$ & $-$ & \SetCell[r=3]{} $3.114$ & $3.079 \pm 0.046$ & \SetCell[r=3]{} $\Psi$ (rad) \\
        & & & & & $1.571 \pm 0.082$ & \\
        & & & & & $0.067 \pm 0.053$ & \\
        \hline
    \end{tblr}
    \caption{On the first column, we show the scalar DM injected parameters in the simulation, with the numerical values on the second column. On the third and fourth columns, we show respectively the scalar DM model best fit and the mean and standard deviation of the full distribution of posteriors. One can notice larger deviation on $\varepsilon$ and $v_\mathrm{DM}$ parameters compared to the other ones. On the right part of the table, we show the GW model on the same DM data, with the corresponding GB parameters. On the fifth and sixth columns, we show the model best fit and the mean and standard deviation of the full distribution of posteriors.}
    \label{tab:DM_DM}
\end{table*}

\begin{figure}[h!]
\centering
\includegraphics[width=0.45\textwidth]{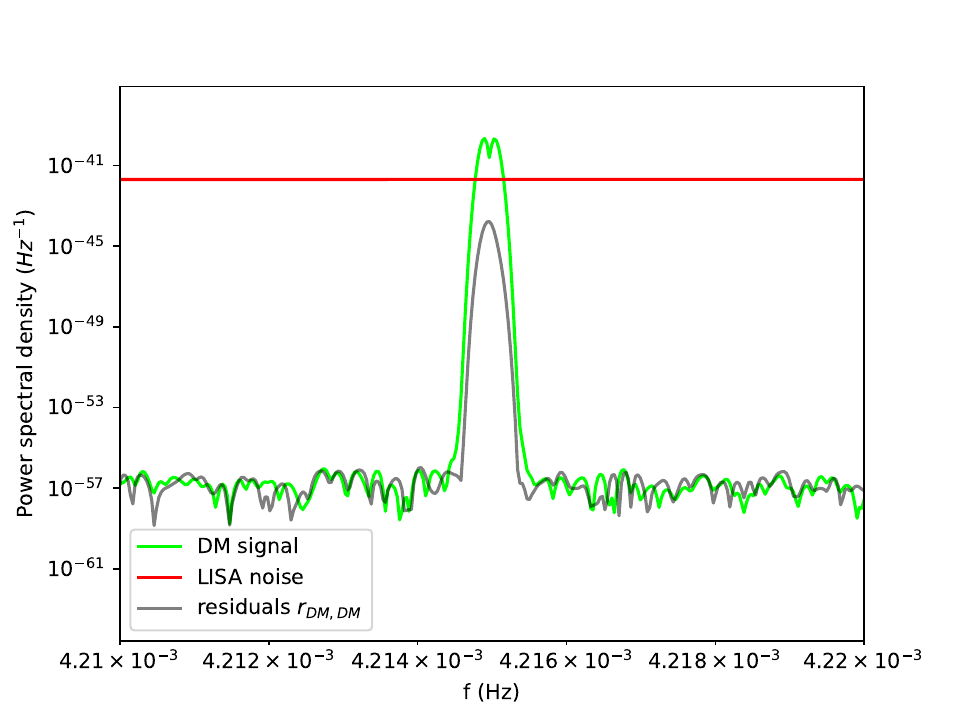}
\caption{DM signal PSD (in lime) compared to LISA noise PSD of TDI $A$ combination (in red) in Fourier domain. The signal is above the noise. We also show the $r_\mathrm{DM,DM}$ residuals power in grey, which are below the noise.}
\label{fig:DM_signal}
\end{figure}

Let us first compute the SNR of such DM data
\begin{subequations}\label{eq:SNR_DM_data}
\begin{align}
    \mathrm{SNR}^\mathrm{DM}_A &\approx 77 \, \\
    \mathrm{SNR}^\mathrm{DM}_E &\approx 65 \, .
\end{align}
\end{subequations}
Note that the SNR differ between the A and E combination due to their different geometric factor, see Eqs.~\eqref{eq:TF_AE}. Although the signal is still larger than the noise (see Fig.~\ref{fig:DM_signal}), the SNR is smaller than in the case of the GW data. This one order of magnitude difference between the DM and GW SNRs come from the difference in transfer functions at low frequency, see Eqs.~\eqref{eq:TDI_X2_DM_low_freq} and \eqref{eq:TDI_X2_GW_low_freq}\footnote{Note that from Fig.~\ref{fig:TF_LISA}, the ratio of the two $X$ transfer functions is $\sim 3$ at $f\sim 4.215$ mHz, but this is because we did not take into account the geometric factors for this plot. However, from Eqs.~\eqref{eq:TDI_X2_DM_low_freq}, \eqref{eq:TDI_X2_GW_low_freq} and \eqref{eq:AET_TDI}, one can easily obtain the low frequency transfer functions of e.g. the TDI $A$ combination. Then, using the true value of DM and GW parameters (i.e $f,\hat e_v$ for DM and $f,\Psi,\imath, \hat k$ for GW), one can compute numerically the various geometric factors $\hat n \cdot \hat e_v$ and $\hat h^\mathrm{SSB}_{ij}\hat n^i\hat n^j$ from the spacecrafts orbits. We find $\mathcal{T}^\mathrm{GW}_A/\mathcal{T}^\mathrm{DM}_A \sim 8.3$ while the ratio of Eqs.~\eqref{eq:SNR_GW_data} and \eqref{eq:SNR_DM_data} gives $\sim 8.2$ for TDI $A$ combination.}.

\paragraph{Fit using a scalar DM model}\label{sec:model_DM_signal_DM}

\begin{figure*}
    \centering
    \includegraphics[width=\textwidth]{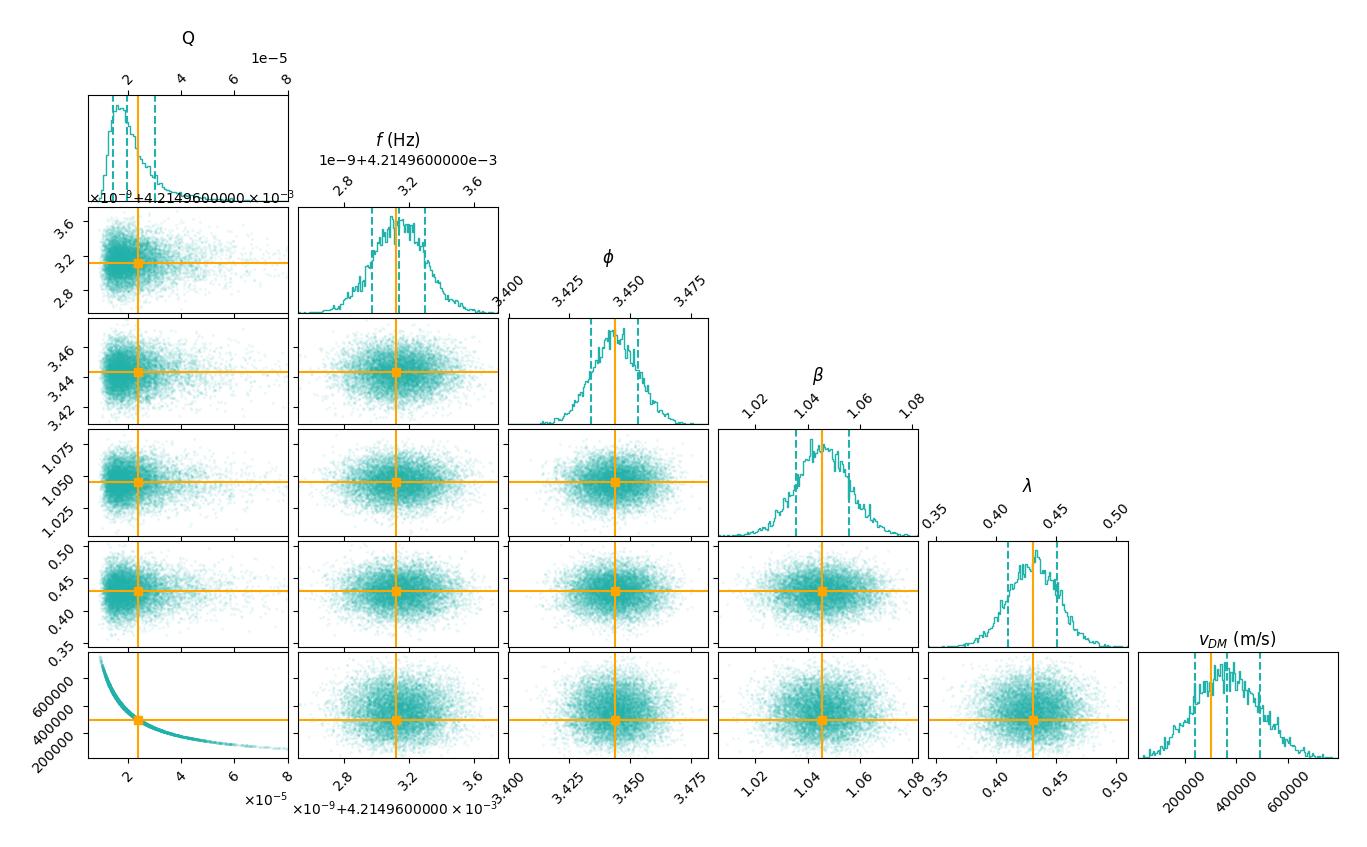}
    \caption{2D posterior distributions of a scalar DM model onto a scalar DM signal with parameters from Tab.~\ref{tab:DM_DM}. The yellow lines indicate the true values of the parameters. As can be noticed from the bottom left joint posterior distribution, the coupling and the wave velocity are highly correlated. The coupling is in units of $10^{-5}$ while the frequency is in units of $10^{-9}$ Hz with an offset of $4.214963$ mHz.}
    \label{fig:cornerplot_DM_DM}
\end{figure*}

We first deal with a scalar DM model on the DM signal presented in the last section. The 2-D posteriors resulting from a fit of such DM model is shown on Fig.~\ref{fig:cornerplot_DM_DM}.

From a technical point of view, instead of sampling the parameters $(Q, v_\mathrm{DM})$ in the fit, we use $(X=Q\times v_\mathrm{DM},v_\mathrm{DM})$ as free parameters. This is motivated because $Q$ and $v_\mathrm{DM}$ are highly correlated (see Fig.~\ref{fig:cornerplot_DM_DM}) while $X$ and $v_\mathrm{DM}$ are marginally correlated. Using these un-correlated parameters in the inference significantly speeds up the convergence of the analysis and reduces sampling errors. In postprocessing, we transform the posteriors on $(X,v_\mathrm{DM})$ into posteriors for the original variables. 

On the left part of Tab.~\ref{tab:DM_DM}, we present the best fit parameters as well as the means and standard deviations of all posterior distributions. As can be noticed from both Tab.~\ref{tab:DM_DM} and Fig.~\ref{fig:cornerplot_DM_DM}, the algorithm correctly recovers the injected value for the fitted parameters (at the 1\% error level) for all parameters, except for the coupling $[Q_M]_d$ and galactic velocity $v_\mathrm{DM}$, whose best fit values differ from the true value by $\sim 20\%$. In addition, the 1-D posterior distribution widths for these two parameters are much larger than for the rest of the parameters ($\sigma_{v_\mathrm{DM}}/v_\mathrm{DM} \sim \sigma_Q/Q \sim 1$ while for the rest of the parameters $\sigma_p/p \ll 1$). The reason is that there exists a strong correlation between both parameters, as can be noticed in the 2-D posterior presented in Fig.~\ref{fig:cornerplot_DM_DM}, which leads to a decrease of sensitivity on each individual parameter when one marginalizes over the other one.  In short, the LISA detector is not sufficiently sensitive to  the DM phase $\varphi$ in order to resolve the propagation phase $\vec k_\phi \cdot \vec x$. Then, from Eq.~\eqref{eq:Doppler_scalar_DM}, the one-link signal behaves schematically as $y \propto Q v_\mathrm{DM}(\cos(\omega_\phi t)-\cos(\omega_\phi (t-L/c)))$ (ignoring the constant phase), i.e $Q$ and $v_\mathrm{DM}$ are maximally correlated. See Sec.~\ref{sec:results} for a more detailed discussion on the effect of this correlation on the determination of $Q$.

Similarly to before, we use the RNR, built from the best fit parameters shown in Table ~\ref{tab:DM_DM}, to assess the goodness of fit:
\begin{subequations}
\begin{align}
    \mathrm{RNR}^\mathrm{DM, DM}_A &\approx 0.35 \, \\
     \mathrm{RNR}^\mathrm{DM, DM}_E &\approx 0.18 \, .
\end{align}
\end{subequations}
Those values are smaller than unity and than the SNR Eq.~\eqref{eq:SNR_DM_data}. As in the GW case above, this suggests the model is efficient in extracting all the information out of the data. This is also visible in Fig.~\ref{fig:DM_signal}, where the residuals are well below the noise power.

\paragraph{Fit using GB model}

We now tackle one of the main goals of this study, i.e we try to answer the following question : can a real scalar DM signal be misinterpreted as a GB ? For this, we try to fit a GW model on simulated data that contains only a DM signal.

\begin{figure}[h!]
    \centering
    \includegraphics[width=0.45\textwidth]{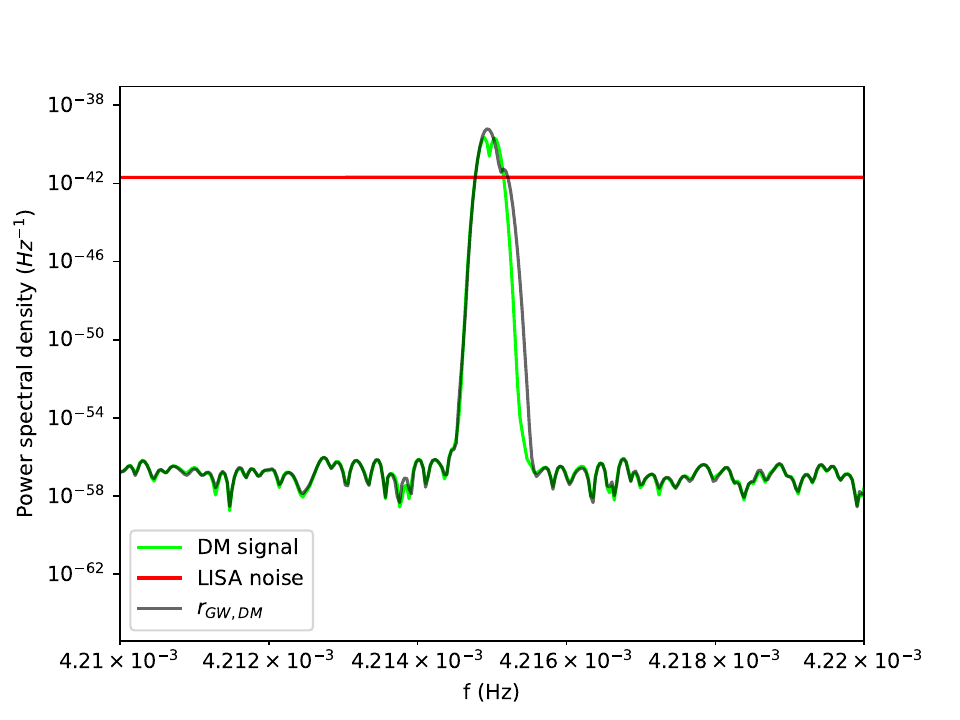}
    \caption{DM signal PSD (in lime) compared to LISA noise PSD of TDI $A$ combination (in red) in Fourier domain (as in Fig.~\ref{fig:DM_signal}). This time, the $r_\mathrm{GW,DM}$ residuals power in grey are close to the signal power and above the noise. This indicates the model fits poorly the data.}
    \label{fig:GW_DM_res}
\end{figure}

The best fit parameters of the GB model are presented in the right part of Tab.~\ref{tab:DM_DM}. We first evaluate the efficiency of the model through residuals and we find
\begin{subequations}
\begin{align}
    \mathrm{RNR}^\mathrm{GW, DM}_A &\approx 51 \, \\
    \mathrm{RNR}^\mathrm{GW, DM}_E &\approx 52 \, .
    \end{align}
\end{subequations}
Those values must be compared with the SNR values of DM data Eq.~\eqref{eq:SNR_DM_data}. One can notice that the SNR and RNR are of the same order of magnitude, and are larger than 1, meaning that the model poorly fits the data. This conclusion is strengthened by Fig.~\ref{fig:GW_DM_res}, where the residuals have a similar power density as the signal. 

\paragraph{Model comparison}

\begin{figure*}
    \centering
    \includegraphics[width=\textwidth]{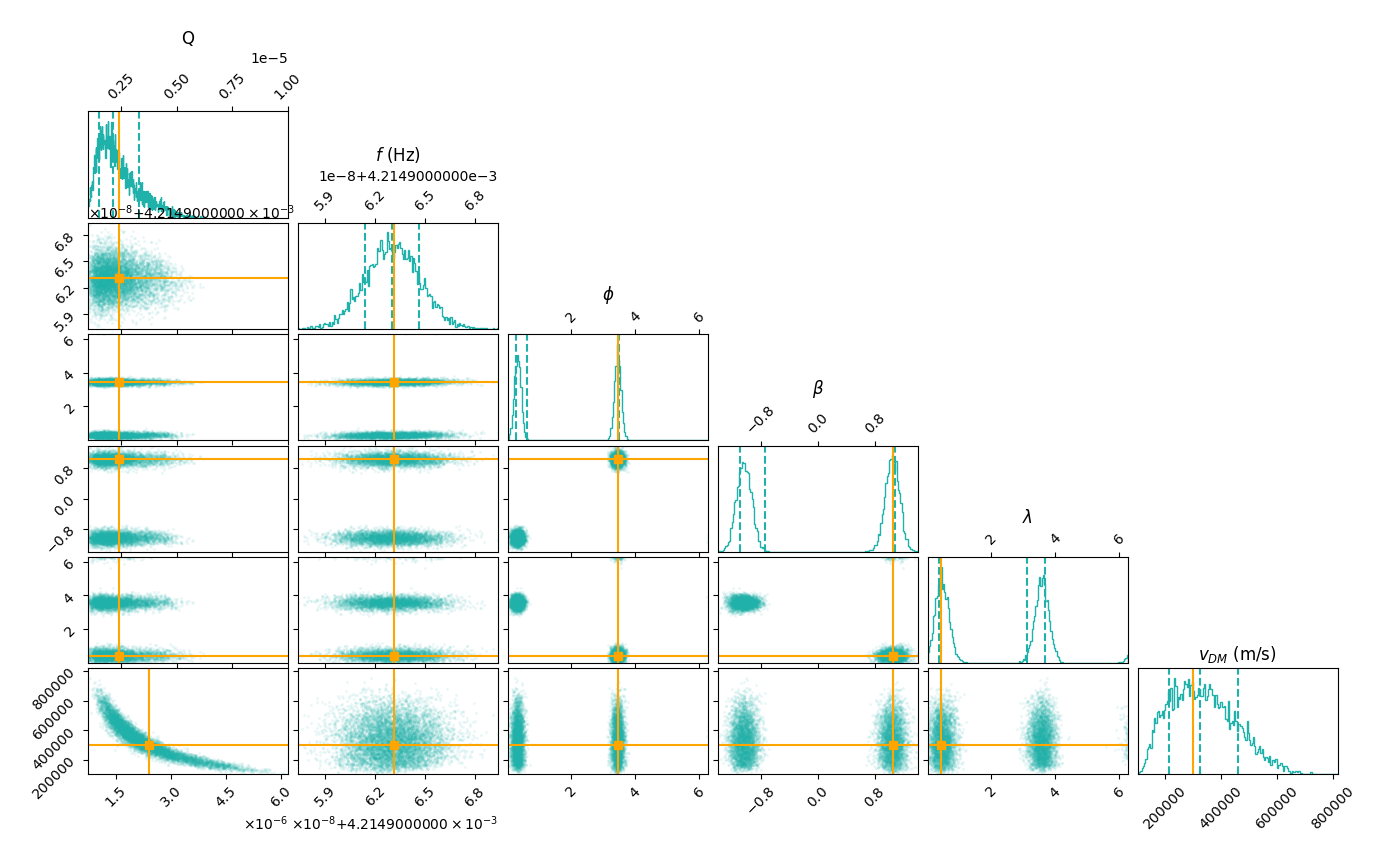}
    \caption{2D posterior distributions of a DM model onto a DM signal with parameters from Tab.~\ref{tab:DM_DM} but with $Q_d \rightarrow Q_d/10$.}
    \label{fig:corner_DN_low_SNR}
\end{figure*}

We now compare the two models by means of the Bayes factor. For the DM signal and both models introduced in the previous sections, we find
\begin{align}
    \log \mathcal{B} &= \mathcal{O}(10^3) \gg 1\, ,
\end{align}
which indicates that the DM model is by far preferred, given the scalar DM signal that was injected in the simulated data. 
To strengthen our findings, we perform another model comparison assuming a scalar DM signal with lower SNR, i.e we generate a dataset containing a DM signal with $Q_d \rightarrow Q_d/10$. We perform a new fit of a DM model, whose posterior distributions are shown in Fig.~\ref{fig:corner_DN_low_SNR}. As before, all the injected parameters are correctly retrieved except the coupling and the velocity, for the same reasons as in the first fit. We then fit a GW model on this dataset and perform a model comparison. We find $\log \mathcal{B} = \mathcal{O}(10)$, i,e the DM model is still preferred over the GW one, with a lower strength of evidence (consistent with the scaling $\mathrm{log}\mathcal{B} \propto \mathrm{SNR}^2$ for Gaussian likelihoods) as compared to the case with larger SNR, but still with decisive preference.
This demonstrates that LISA will be able to make the distinction between a DM signal (with a Compton frequency within the LISA band and with an interaction parameterized by Eq.~\eqref{dilaton_lagrangian}) and a GB if using more than one year of data and if the signal is larger than the LISA noise. This is one of the main results of this paper, which was not explored previously. 

This means that all the previous studies (see e.g. \cite{Yu23, Morisaki21, Morisaki19, Pierce18, Yao24} which derived the sensitivity of LISA to ULDM couplings inducing oscillating accelerations on test masses, are valid in the sense that with realistic orbits and at least one year of data, a scalar DM has a singular signature on the detector, different from the ones of GB, and therefore it can be identified.

\section{Sensitivity of LISA to ULDM couplings}\label{sec:results}

In this section, we estimate the upper limit sensitivity of LISA to ULDM couplings, which in our simulation is parameterized generically by $Q$. 

Since we are interested in inferring an estimate of an upper limit that LISA would set on such a coupling parameter, we will work under the assumption that there is no  signal in the data. This is a major difference compared to the results presented in the previous section where either a GW or a DM signal was injected. The fact that no signal is present in the data impacts significantly the correlation between $Q$ and $v_\mathrm{DM}$ in the posterior distribution. Indeed, when a signal is present in the data, we have shown in Sec.~\ref{sec:fit_DM} that these two parameters are highly correlated in the posterior, see Fig.~\ref{fig:cornerplot_DM_DM} and that the posterior on $Q$ depends highly on the prior on $v_\mathrm{DM}$. We will show here that these conclusions do not apply when no signal is detected. 

\begin{figure*}
    \centering
    \includegraphics[width=0.49\textwidth]{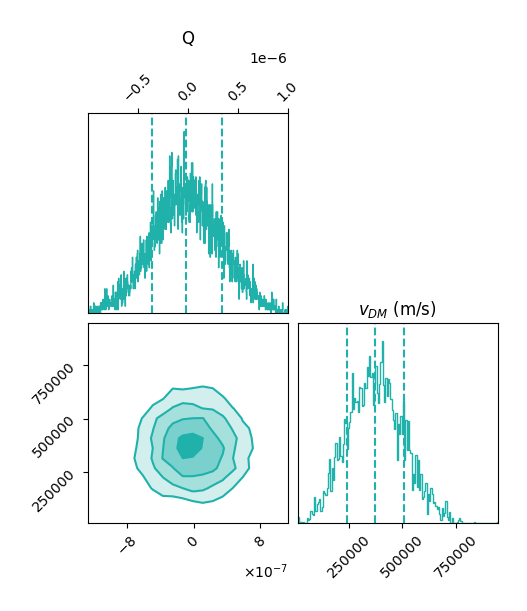}
    \includegraphics[width=0.49\textwidth]{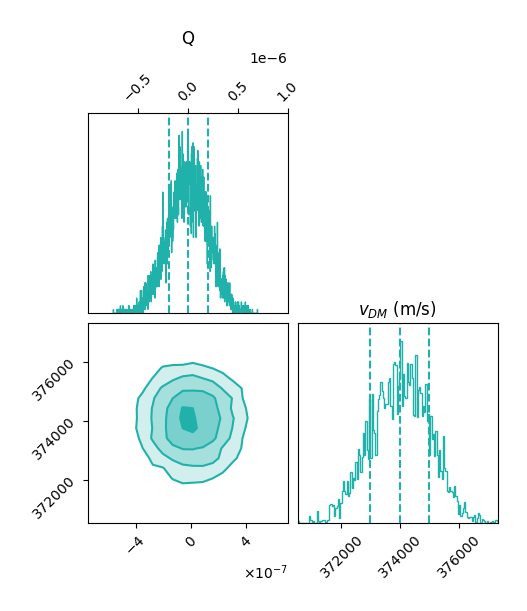}
    \caption{2D posterior distributions of a DM model onto a dataset which contains no  signal, i.e where the coupling $Q \rightarrow 0$. Left: the prior on the DM velocity is the standard velocity distribution from Eq.~\eqref{DM_vel_distrib}. Right: the prior on $v_\mathrm{DM}$ is much tighter, i.e we decreased the standard deviation by a factor $\sim 100$. One can notice that this does not change the posterior distribution of $Q$, in particular its standard deviation, by a factor $ < 2$.}
    \label{fig:corner_no_sig}
\end{figure*}

First, we will show that $Q$ and $v_\mathrm{DM}$ are not correlated by performing a Bayesian inference on a mock dataset which contains no DM signal (i.e $Q\rightarrow 0$). The result of such an analysis is presented on the left panel of Fig.~\eqref{fig:corner_no_sig}. One can notice that, unlike Fig.~\eqref{fig:cornerplot_DM_DM}, the correlation between $Q$ and $v_\mathrm{DM}$ disappears when $Q\rightarrow 0$.

 The reason for this behaviour can be explained analytically. As discussed in Sec.~\ref{sec:model_DM_signal_DM}, LISA data will be sensitive to the amplitude of the signal $X=Q\times v_\mathrm{DM}$ and will be insensitive to the phase of the signal and therefore insensitive to $v_\mathrm{DM}$ alone. If we stay general for now and consider that the posterior of $X$ follows a Gaussian distribution with mean $\mu_X$, we can expand the ratio $X/v_\mathrm{DM}$ at linear order to find that the variance of $Q$ is 
 \begin{subequations}
 \begin{align}
     \sigma^2_Q &= \frac{\mu^2_X}{\mu^2_v}\left(\frac{\sigma^2_X}{\mu^2_X}+\frac{\sigma^2_v}{\mu^2_v}\right) \, ,
 \end{align}
 when there is no correlation between $X$ and $v_\mathrm{DM}$. Similarly the correlation between $Q$ and $v_\mathrm{DM}$ writes at first order as
\begin{equation}
    \sigma_{Qv} = -\mu_X^2 \frac{\sigma_v^2}{\mu_v^2} \, .
\end{equation}
\end{subequations}
Clearly, when there is no signal in the data, i.e when $\mu_X \rightarrow 0$, the correlation between $Q$ and $v_\mathrm{DM}$ in the posterior disappears, and we find that the width of the marginalized posterior on $Q$ is $\sigma_Q = \sigma_X/\mu_v$, i.e it becomes independent of the width of the velocity distribution $\sigma_v$.
 
To further validate this result and to clearly demonstrate that the width of the $v_\mathrm{DM}$ distribution does not impact significantly the width of the posterior on $Q$, we present in Fig.~\ref{fig:corner_no_sig} (right panel) the result of a fit where a very tight prior on $v_\mathrm{DM}$ is used. The variance of the posterior on $Q$ is similar to the one obtained  using a realistic (wider) prior on $v_\mathrm{DM}$ (see left panel). 

This demonstrates that, when the data contains no DM signal, the sensitivity of the detector to $Q$ is the same whether the DM velocity is a free parameter or fixed to its mean value. In the following, we will then derive the sensitivity of LISA to ULDM couplings, assuming $v_\mathrm{DM}$ fixed, as it was done in previous works, see e.g. \cite{Yu23}.

\begin{figure*}
    \centering
    \includegraphics[width=\textwidth]{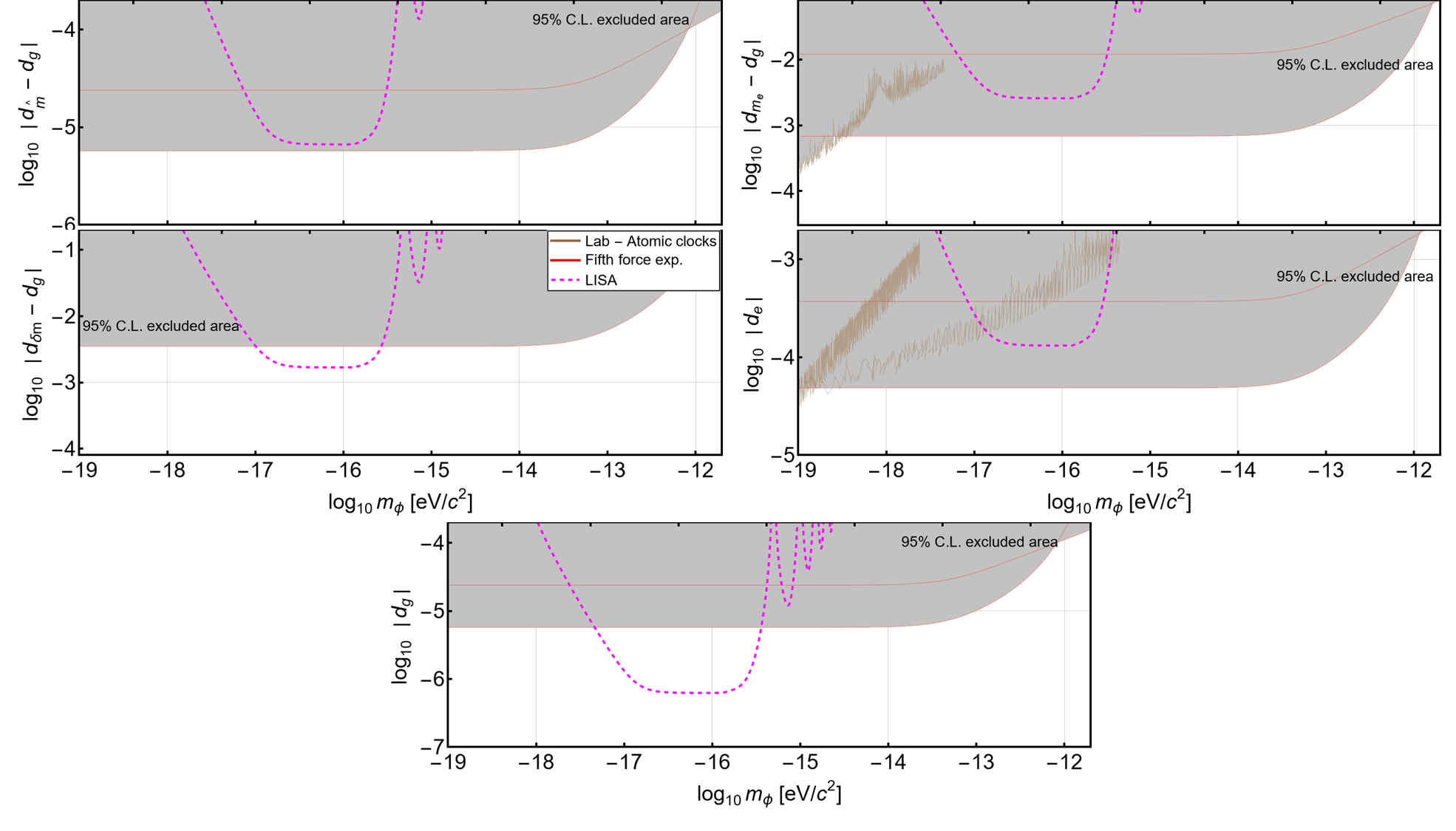}
    \caption{Current constraints on all the dilatonic couplings of interest in this paper : $d_{\hat m}-d_g$ (first line left), $d_{\delta\hat m}-d_g$ (second line left), $d_{m_e}-d_g$ (first line right), $d_e$ (second line right), $d_g$ (bottom) from \cite{Microscope22,Wagner12,Hees16,Kennedy20}, with 95\% confidence level (shown in light grey background). All existing constraints are shown in solid lines. The expected sensitivity of LISA is shown in pink dashed lines (with 68\% detection threshold).}
    \label{fig:full_dilaton_constraints}
\end{figure*}
\begin{figure}
    \centering
    \includegraphics[width=0.5\textwidth]{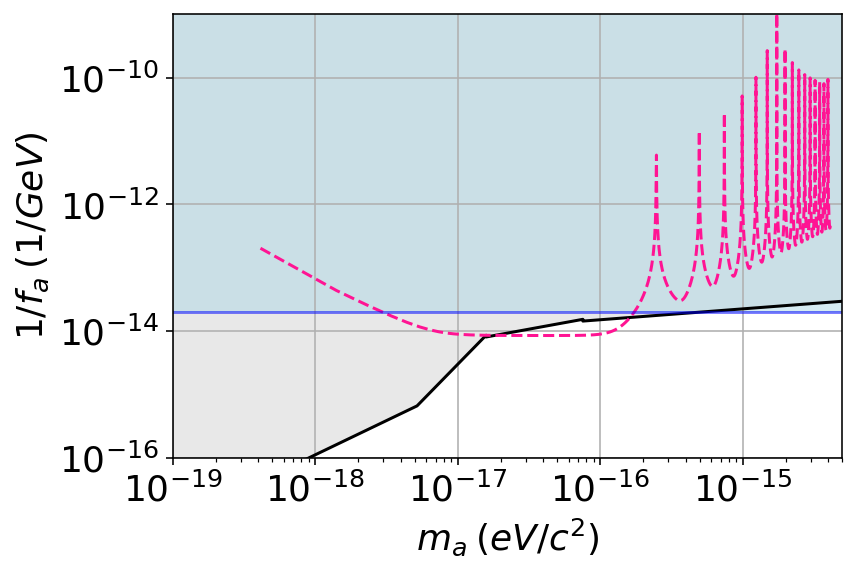}
    \caption{In grey background are shown the current lab constraints on $1/f_a$ axion coupling \cite{Beam_EDM,HfF,Rb_quartz,nEDM, Gue25} (the constraint from \cite{Rb_quartz} has been rescaled for consistent value of local DM energy density). In blue is shown the Earth bound constrain from \cite{Banerjee25, Gan25} which corresponds to a region where a phase transition of the axion field inside Earth would modify the field amplitude in its vicinity. The sensitivity estimate of LISA is shown in pink dashed lines.}
    \label{fig:constraints_fa}
\end{figure}

\subsection{Dilaton couplings}
In this section, we will estimate the sensitivity of LISA to the dilaton coupling. The sensitivity will be defined as the coupling amplitude for which the signal-to-noise ratio (SNR) is one.

Using Eq.~\eqref{eq:Doppler_scalar_DM}, the one-sided power spectral density (PSD) of the (monochromatic) signal we are interested in reads
\begin{align}\label{eq:signal_PSD}
    S_{A}(f) &= \left(\frac{\sqrt{8\pi G \rho_\mathrm{DM}}v_\mathrm{DM}Q_d}{2\pi f_\phi c^2}\right)^2 \times \mathcal{T}^2_{A} \times \, \nonumber \\
    & \frac{1}{2\pi^2 T_\mathrm{obs}}\left(\frac{\sin (\pi(f-f_\phi)T_\mathrm{obs})}{f-f_\phi}\right)^2\, ,
\end{align}
and similarly for $E$, where $f$ is the Fourier frequency, where we used a square window of duration $T_\mathrm{obs}$ for simplicity and where $\mathcal{T}_{A}$ is the DM transfer function of $A$ given in Ap.~\ref{ap:TF}. As pointed out in Eq.~\eqref{eq:TDI_A_noise_PSD}, these two TDI combinations share the same noise PSD, $N_A=N_E\equiv N_{A,E}$. At SNR=1, the sensitivity of one given TDI combination (say $A$) is obtained by equating Eq.~\eqref{eq:signal_PSD} with Eq.~\eqref{eq:TDI_A_noise_PSD}. However, in our case, we are interested in the potential of DM detection using the optimum data weighting which combines $A,E,T$ combinations, and in this case, the total SNR becomes \cite{Prince02,Yu23}
\begin{align}
   \mathrm{SNR} =  \frac{S_A}{N_A} + \frac{S_E}{N_E} + \frac{S_T}{N_T}  \, .
\end{align}
However, as pointed out in \cite{Yu23}, the sensitivity of the $T$ combination to scalar DM  is several orders of magnitude smaller than $A,E$ such that we will only consider the latter two in the following.
Using the SNR=1 criteria, we find the sensitivity of LISA to the coupling $Q$ to be (in the limit where $f \rightarrow f_\phi$), 
\begin{subequations}\label{eq:coupling_no_corr}
\begin{align}
    Q_d(f) &= \frac{2\pi f c^2 }{\sqrt{8\pi G \rho_\mathrm{DM}} v_\mathrm{DM}}\sqrt{\frac{N_{A,E}(f)}{\mathcal{T}^2_{A,E} T_\mathrm{obs}}} \label{eq:coupling_low_freq}\, ,
\end{align}
where $\mathcal{T}^2_{A,E}(f)=\mathcal{T}^2_{A}+\mathcal{T}^2_{E}$ is the joint transfer function of $A$ and $E$ (see Ap.~\ref{ap:TF}).

This equation works well when the DM field can effectively be modelled as a monochromatic wave, i.e when $T_\mathrm{obs} \ll \tau(f)$, where $\tau(f) \sim 10^6/f$ is its coherence time \cite{Derevianko18}. 
Note that a correction factor to the sensitivity arises due to the stochastic nature of the amplitude of the field \cite{Centers21} which induces a loss in signal of $\sim 1.5$.
When the integration time is much longer than the coherence time of the field, i.e $T_\mathrm{obs} \gg \tau(f)$, the signal searched for is no longer coherent, i.e. it should be modeled as a sum of several stochastic harmonics, see \cite{foster:2018aa}, and therefore Eq.~\eqref{eq:signal_PSD} is not correct anymore. In this case, one needs to cut the dataset in subsets with duration smaller than $\tau(f)$ and search for a coherent signal in each of these blocks of data. In such a case, the experimental sensitivity to the coupling is reduced and becomes \cite{Budker14}
\begin{align}
    Q_d(f) &= \frac{2\pi f c^2}{\sqrt{8\pi G \rho_\mathrm{DM}} v_\mathrm{DM}}\sqrt{\frac{N_{A,E}(f)}{\mathcal{T}^2_{A,E}\sqrt{T_\mathrm{obs}\tau(f)}}} \,  \label{eq:coupling_high_freq}.
\end{align}
\end{subequations}

Since $Q_d = d_g+\sum_i Q^\mathrm{TM}_{i} \: d_i$ (from Eq.~\eqref{partial_dil_mass_charge}), where $Q^\mathrm{TM}_i$ is the partial dilatonic charge of the test mass, the sensitivity of LISA to the coupling $d_i$ reads (assuming only one non-zero coupling at a time)
\begin{align}\label{eq:di_varepsilon}
    d_i &= \frac{Q_d}{Q^\mathrm{TM}_{i}} \, ,
\end{align}
and $d_g = Q_d$ because of the first term in Eq.~\eqref{partial_dil_mass_charge}. LISA will use test masses formed from 73\% gold and 27\% platinum \cite{Diaz12}. Using Eqs.~\eqref{dilatonic_partial_mass_charge}, we compute the respective partial dilatonic charges and we find \begin{subequations}
\begin{align}
    Q^\mathrm{TM}_e &= 4.3 \times 10^{-3}\, ,\\
    Q^\mathrm{TM}_{m_e} &= 2.2\times 10^{-4}\,, \\
    Q^\mathrm{TM}_{\hat m} &= 8.53 \times 10^{-2}\,, \\
    Q^\mathrm{TM}_{\delta m} &= 3.37 \times 10^{-4}\, .
\end{align}
\end{subequations}

Using Eqs.~\eqref{eq:di_varepsilon} together with \eqref{eq:coupling_low_freq} and \eqref{eq:coupling_high_freq}, we can now derive the sensitivity curves. 

In Fig.~\ref{fig:full_dilaton_constraints}, we show in pink the sensitivity of LISA to the different dilatonic $d_i$ couplings. Following the decomposition from Eq.~(\ref{partial_dil_mass_charge}), we present the sensitivity of LISA to $d_g$, $d_e$ (which depends on the charge $Q^\mathrm{TM}_{e}$ of the test mass) and to all the other couplings of the form $d_i-d_g$ (with $i=m_e,\hat m, \delta m$) which depend on one partial charge of the test mass only ($Q^\mathrm{TM}_i$).

One can notice that LISA will reach unconstrained regions of the parameter space, in particular on the $d_g$ coupling and also marginally on the $d_{\hat m} -d_g$ couplings. More precisely, LISA will improve fifth force constraints by an order of magnitude for $d_g$ and by a factor $\sim 2$ over a mass range of one order of magnitude ($\sim 10^{-17} \: \mathrm{eV}-10^{-16} \: \mathrm{eV}$) for the other. These estimates are consistent with the literature, e.g. \cite{Yu23}, the only difference is the additional loss factor $\sim 1.5$ due to the stochastic nature of DM (see discussion above). 

\subsection{Axion-gluon coupling}

Similarly as for the dilatonic couplings, LISA will be able to probe the axion-gluon coupling through Doppler effects, see Eq.~\eqref{eq:Doppler_pseudoscalar_DM}.  
The sensitivity on $f_a$ is computed using the same method as the one presented in the previous section, which leads to
\begin{subequations}
\begin{align}
    \left(\frac{E_P}{f_a}\right)^2 &= \frac{(2\pi f)^2 c^3}{8\pi G \rho_\mathrm{DM} v_\mathrm{DM}}\sqrt{\frac{N_{A,E}(2f)}{\left|\mathcal{T}_{A,E}\right|^2T_\mathrm{obs}}} \,\\
    &\mathrm{or} \, \nonumber \\
   \left(\frac{E_P}{f_a}\right)^2 &= \frac{(2\pi f)^2 c^3}{8\pi G \rho_\mathrm{DM}v_\mathrm{DM}}\sqrt{\frac{N_{A,E}(2f)}{\left|\mathcal{T}_{A,E}\right|^2\sqrt{T_\mathrm{obs}\tau(f)}}}\, ,
\end{align}
\end{subequations}
respectively for $T_\mathrm{obs} \ll \tau(f)$ and $T_\mathrm{obs} \gg \tau(f)$, where we took into account the different normalization amplitude of the signal (from Eq.~\eqref{eq:Doppler_pseudoscalar_DM}) and that it oscillates at twice the axion frequency.

In Fig.~\ref{fig:constraints_fa}, we show in pink the sensitivity of LISA to $1/f_a$ coupling.  At axion masses between $10^{-17}$ and $10^{-14}$ eV, LISA would be competitive with current best laboratory constraints set by MICROSCOPE \cite{Gue25}, improving them by a factor $\mathcal{O}(1)$. Note that this coupling can also be constrained by the observation of astrophysical objects such as white dwarfs or neutron stars, as it modifies the nucleon effective mass inside the body \cite{Balkin23,Balkin24, Gomez24,Kumamoto25}.

\section{Discussion}\label{sec:discussion}

Through their couplings with various SM sectors, some scalar ULDM candidates produce an quasi monochromatic oscillation of the motion of freely falling test masses which will generate a Doppler shift in light beams exchanged between them. For this reason, it has been suggested to use space-based GW detectors such as LISA to also search for ULDM. Since GW from quasi monochromatic sources (e.g. galactic binaries) produce a very similar effect, a natural question that arises is whether the data can efficiently break the possible degeneracy between the two signals.
In this paper, we answer this question by using Bayesian inference and one year of  LISA data using realistic orbits. We begin by deriving a fast likelihood modeling of a scalar ULDM based on \cite{Cornish07}. 
Using this, we find that LISA will surely be able to make the difference between scalar ULDM and a GW from a galactic binary. Indeed, using a model selection tool, namely the Bayes factor, we find that in the case of a mock scalar ULDM signal, the ULDM model is preferred over the GW one by a factor $\sim 10^3$, while the mock GW data is much better fitted by a galactic binary GW model, compared to a scalar ULDM (the Bayes factor reaches $\sim 10^5$). Then, we derive the sensitivity of LISA to both dilaton and axion couplings, and we show that it will be able to probe unconstrained regions of the parameter spaces, compared to existing laboratory bounds.
Our conclusion that LISA will be able to fully decorrelate a ULDM signal from any individual galactic binary signal is, rigorously speaking, only true for a single GB and ULDM. It remains to be seen if it also holds in the presence of the full background of thousands of GB and other GW sources of deterministic or stochastic nature expected in LISA. Studying that question is the main perspective for continuing the work presented here.

\begin{acknowledgments}
This work was supported by the Programme National GRAM of CNRS/INSU with INP and IN2P3 cofunded by CNES. IFAE is partially funded by the CERCA program of the Generalitat de Catalunya. J.G. is funded by the grant CNS2023-143767, funded by MICIU/AEI/10.13039/501100011033 and by
European Union NextGenerationEU/PRTR. 
\end{acknowledgments}
\newpage
\appendix

\begin{widetext}
\section{Velocity perturbation of a test mass}\label{ap:delta_xA}

In this appendix, we compute the perturbed velocity of the test mass B from the acceleration Eq.~\eqref{EP_viol_acc_rm}.
One can treat the position $\vec x(t)$ in Eq.~\eqref{EP_viol_acc_rm} as the unperturbed position $\vec x_0(t)$ at leading order and write it as $\vec x(t) = \vec x_0(t) + \delta \vec x_B(t)$, where the latter is $\mathcal{O}(\sqrt{\rho_\mathrm{DM}}Q^B_d)$. Considering that $\vec x(t) = \vec x_\mathrm{AU} \cos(\omega_E t)$ at leading order, where $|\vec x_\mathrm{AU}| \sim 1.5 \times 10^{11}$m is one astronomical unit distance and $\omega_E$ is the Earth rotation frequency around the Sun, one can integrate one this expression, which leads to the oscillation of the position of the test mass\footnote{To make this integration, one needs to expand the $\sin(\omega_\phi t - \vec k_\phi \cdot \vec x +\Phi)$ in power of $\vec k_\phi \cdot \vec x_\mathrm{AU} \cos(\omega_E t)\ll 1$, integrate each term of the series and then reconstruct the cosine. We also considered $\omega_E \ll \omega_\phi$.}
\begin{align}
    &\delta \vec v_A(t,t_0,\vec x) = \frac{-\sqrt{8 \pi G \rho_\mathrm{DM}}\vec v_\mathrm{DM}}{\omega_\phi c} Q^A_d \left[\cos(\omega_\phi t - \vec k_\phi \cdot \vec x_\mathrm{AU}\cos(\omega_E t) +\Phi)-\cos(\omega_\phi t_0 - \vec k_\phi \cdot \vec x_\mathrm{AU}\cos(\omega_E t_0) +\Phi)\right] \,  .
\end{align}
The last term has no time dependence and therefore will have no impact on the analysis. The slow oscillation at $\omega_E$ will produce side bands in the Fourier signal at frequencies $\omega_\phi\pm \omega_E$. However, note that for our analysis, we assumed a time of integration of one year $T_\mathrm{obs}=2\pi/\omega_E$, such that the size of one Fourier bin is exactly $f_E=\omega_E/2\pi$. Therefore, these side bands might not be visible but will most likely enlarge the signal at $\omega_\phi$.

\section{Transfer functions}\label{ap:TF}

In this appendix, we compute the second generation TDI $X_2$ combination transfer functions for scalar DM and monochromatic GW.

\subsection{Scalar ultralight dark matter}

In the constant armlength limit, we use Eq.~\eqref{eq:TDI_X2_approx} together with Eq.~\eqref{eq:TF_DM} such that the scalar DM transfer function of $X_2$ is
\begin{subequations}
\begin{align}
&\mathcal{T}^\mathrm{DM}_X(\omega) =-4\sin \left(\frac{\omega L}{c}\right)\sin \left(\frac{2\omega L}{c}\right)\Re\left[e^{-3i\omega L/c} \left(\mathcal{T}^\mathrm{DM}_{13} -\mathcal T^\mathrm{DM}_{12} + e^{-i\omega L/c}\left(\mathcal T^\mathrm{DM}_{31}-\mathcal T^\mathrm{DM}_{21}\right)\right)\right]\, \\
&= -4\sin \left(\frac{\omega L}{c}\right)\sin \left(\frac{2\omega L}{c}\right)\Re\left[e^{-3i\omega L/c}e^{-i\vec k \cdot \vec x_1}\left(1+e^{-2i\omega L/c}\right)\left(\hat n_{13}-\hat n_{12}\right)\cdot \hat e_v -\right.\,\nonumber \\
&\left.2e^{-4i\omega L/c}\left(\hat n_{13}\cdot \hat e_v e^{-i\vec k \cdot \vec x_3}-\hat n_{12}\cdot \hat e_v e^{-i\vec k \cdot \vec x_2}\right)\right]\, \\
&= -8\sin \left(\frac{\omega L}{c}\right)\sin \left(\frac{2\omega L}{c}\right)\Re\left[e^{-4i\omega L/c}\left(\left(\hat n_{13}-\hat n_{12}\right)\cdot \hat e_v \cos\left(\frac{\omega L}{c}\right)e^{-i\vec k \cdot \vec x_1 }-\right.\right.\,\nonumber \\
&\left.\left.\left(\hat n_{13} e^{-i\vec k \cdot \vec x_3}-\hat n_{12} e^{-i\vec k \cdot \vec x_2}\right)\cdot \hat e_v\right)\right] \, \\
&=-8\sin \left(\frac{\omega L}{c}\right)\sin \left(\frac{2\omega L}{c}\right)\Re\left[e^{-i\left(4\omega L/c+\vec k \cdot \vec x_1\right)}\left(\hat n_{13}\cdot \hat e_v\left(\cos\left(\frac{\omega L}{c}\right)-1+ikL \hat n_{13}\cdot \hat e_v\right)-\right.\right.\,\nonumber\\
&\left.\left.\hat n_{12}\cdot \hat e_v\left(\cos\left(\frac{\omega L}{c}\right)-1+ikL \hat n_{12}\cdot \hat e_v\right)\right)\right]\label{eq:TF_X2_DM_full}\,,
\end{align}
\end{subequations}
where we used $\vec x_2 = \vec x_1 - L \hat n_{12}$ and $\vec x_3 = \vec x_1 - L\hat n_{13}$ and we took the first order expansion of the exponent $\propto kL \ll 1$ for all frequencies of interest (for example, at $f=1$ Hz, which is the maximum frequency of the LISA band, $kL \sim 0.05$, with the galactic velocity $v_\mathrm{DM} = 10^{-3} \: c$). The amplitude of the transfer function is then
\begin{align}
&\left|\mathcal{T}^\mathrm{DM}_X(\omega)\right| = 16\left|\sin \left(\frac{\omega L}{c}\right)\sin \left(\frac{2\omega L}{c}\right)\right| \sqrt{\left(\hat n_{23}\cdot \hat e_v\right)^2\sin^4\left(\frac{\omega L}{2c}\right)+\left((\hat n_{13}\cdot \hat e_v)^2-(\hat n_{12}\cdot \hat e_v)^2\right)^2\left(\frac{\omega L|\vec v_\mathrm{DM}|}{2c^2}\right)^2} \, ,
\end{align}
where we used $\hat n_{13}-\hat n_{12}=\hat n_{23}$. As it can be noticed from the above equation, the first term dominates as long as $(\omega L/c)^2 > (|\vec v_\mathrm{DM}|/c)^2$ (and $\overline{(\hat n_{23} \cdot \hat e_v)^2} \neq 0$), i.e for all frequencies $f \geq 5 \times 10^{-5}$ Hz, which fully contains the LISA band. Then, we can neglect the second term and we recover the transfer function Eq.~\eqref{eq:amp_TF_X2}.

For our analysis, we use the $A$ and $E$ combinations which are defined in Eq.~\eqref{eq:AET_TDI}.
As mentioned in the main text, the transfer function of the $Y$ and $Z$ combinations can be obtained from the $X$ one respectively by the $\hat n_{23} \rightarrow \hat n_{31}$ and $\vec x_1 \rightarrow \vec x_2$ substitutions for $Y $ and $\hat n_{23} \rightarrow \hat n_{12}$ and $\vec x_1 \rightarrow \vec x_3$ changes for $Z$
Then, we find that the transfer functions of $A$ and $E$ read
\begin{subequations}\label{eq:TF_AE}
\begin{align}
    &\left|\mathcal{T}^\mathrm{DM}_A(\omega)\right| = \frac{16}{\sqrt{2}}\left|\sin \left(\frac{\omega L}{c}\right)\sin \left(\frac{2\omega L}{c}\right)(\hat n_{12}-\hat n_{23})\cdot \hat e_v\right|\sin^2\left(\frac{\omega L}{2c}\right) \, \\
    &\left|\mathcal{T}^\mathrm{DM}_E(\omega)\right| = \frac{16\sqrt{3}}{\sqrt{2}}\left|\sin\left(\frac{\omega L}{c}\right)\sin \left(\frac{2\omega L}{c}\right)\hat n_{13}\cdot \hat e_v\right|\sin^2\left(\frac{\omega L}{2c}\right) \, .
\end{align}
\end{subequations}
Then, the joint transfer function of the 2 combinations is the sum of the squares, i.e 
\begin{align}
    &\left|\mathcal{T}^\mathrm{DM}_{A,E}(\omega)\right| = \frac{16}{\sqrt{2}}\sqrt{\mu_\mathrm{DM}}\left|\sin \left(\frac{\omega L}{c}\right)\sin \left(\frac{2\omega L}{c}\right)\right|\sin^2\left(\frac{\omega L}{2c}\right) \, ,
\end{align}
with 
\begin{align}
    \mu_\mathrm{DM} &= \left(\sqrt{3}\hat n_{13}\cdot \hat e_v\right)^2+\left((\hat n_{12}-n_{23})\cdot \hat e_v\right)^2 \, .
\end{align}
Numerically, we find that $\mu_\mathrm{DM} \sim 4.7$, neglecting the oscillation at Earth's orbital frequency, far below the LISA band.

\subsection{Monochromatic gravitational waves}

Using Eqs.~\eqref{eq:TF_GW} and \eqref{eq:TDI_X2_approx}, the transfer function of the monochromatic GW reads 
\begin{subequations}
\begin{align}\label{eq:TDI2_GW_TF}
    &\mathcal{T}^\mathrm{GW}_X(\omega) = 2\sin \left(\frac{\omega L}{c}\right)\sin \left(\frac{2\omega L}{c}\right) \Re\left[\hat h^\mathrm{SSB}_{ij} e^{-i\left(\frac{3\omega L}{c}+\vec k \cdot \vec x_1\right)} \: \times \right.\,\\
    &\left.\sum_{\ell=2,3} w_\ell \frac{\hat n^i_{1\ell} \hat n^j_{1\ell}}{1-\left(\hat n_{1\ell} \cdot \hat k  \right)^2}\left((1+\hat n_{1\ell} \cdot \hat k)\left(1-e^{-i\frac{\omega L}{c}(1-\hat n_{1\ell} \cdot \hat k)}\right)-e^{-i\frac{2\omega L}{c}}(1-\hat n_{1\ell} \cdot \hat k)\left(1-e^{i\frac{\omega L}{c}(1+\hat n_{1\ell} \cdot \hat k)}\right)\right)\right]\, \nonumber \\
    &=2\sin \left(\frac{\omega L}{c}\right)\sin \left(\frac{2\omega L}{c}\right) \Re\left[\hat h^\mathrm{SSB}_{ij} e^{-i\left(\frac{5\omega L}{c}+\vec k \cdot \vec x_1 \right)} \: \times \right.\,\\
    &\left.\sum_{\ell=2,3} w_\ell \frac{\hat n^i_{1\ell} \hat n^j_{1\ell}}{1-\left(\hat n_{1\ell} \cdot \hat k  \right)^2}\left((\hat n_{1\ell} \cdot \hat k)\left(1-2e^{\frac{i \omega L}{c}(1+\hat n_{1\ell} \cdot \hat k)}+e^{\frac{2i\omega L}{c}}\right)+2ie^{\frac{i\omega L}{c}}\sin\left(\frac{\omega L}{c}\right)\right)\right] \, \nonumber,
\end{align}
where $w_\ell = \pm 1$ respectively for $\ell=3,2$. One can simplify this expression when $\omega L/c \ll 2\pi$ to find the amplitude of the transfer function
\begin{align}
    |\mathcal{T}^\mathrm{GW}_X(\omega)| &= 8\left(\frac{\omega L}{c}\right)^3\left|\hat h^\mathrm{SSB}_{ij} \left(\hat n^i_{13}\hat n^j_{13}-\hat n^i_{12}\hat n^j_{12}\right)\right| \, .
\end{align}
\end{subequations}
\end{widetext}

\section{Fast likelihood for scalar ULDM}\label{ap:fastDM_fastGB}

In this appendix, we derive (in time domain) the slowly oscillating component of TDI $X$ combination of a scalar ULDM signal. 

We will make the detailed calculations for a (pure scalar) DM signal.
Let us start again with the one-link response function induced by scalar DM Eq.~\eqref{eq:Doppler_scalar_DM} at time of reception of the photon of the receiver $r$ $t_r$, when it is emitted at time $t_e=t_r-L/c$ by emitter $e$ 
\begin{subequations}\label{eq:Doppler_DM}
\begin{align}
&y^\mathrm{DM}_{re}(t_r) = \left(\hat n_\mathrm{re} \cdot \hat e_v\right)\frac{\sqrt{8 \pi G \rho_\mathrm{DM}}v_\mathrm{DM}Q_d}{\omega_\phi c^2}\Re\left[e^{i\phi_r}- e^{i\phi_e}\right] \,\\
&=\left(\hat n_\mathrm{re} \cdot \hat e_v\right)\frac{\sqrt{8 \pi G \rho_\mathrm{DM}}v_\mathrm{DM}Q_d}{\omega_\phi c^2}\Re\left[e^{i\phi_r}\left(1- e^{-2i\alpha^\mathrm{DM}_{re}}\right)\right]\, \\
&=2\left(\hat n_\mathrm{re} \cdot \hat e_v\right)\frac{\sqrt{8 \pi G \rho_\mathrm{DM}}v_\mathrm{DM}Q_d}{\omega_\phi c^2}\sin(\alpha^\mathrm{DM}_{re})\Re\left[ie^{i\phi_r}e^{-i\alpha^\mathrm{DM}_{re}}\right]\, \\
&\equiv Y_{re}(t_r) e^{i\phi_r} \, ,
\end{align}
\end{subequations}
where
\begin{subequations}
\begin{align}
\phi_r  &= \omega_\phi t_r- \vec k_\phi \cdot \vec x_r(t_r) + \Phi\, \\
\phi_e  &= \omega_\phi t_e  -\vec k_\phi \cdot \vec x_e(t_e)  + \Phi  \, ,
\end{align}
and therefore
\begin{align}
    &2\alpha^\mathrm{DM}_{re}(t_r) = \phi_r-\phi_e \,\\
    &= \omega_\phi\left((t_r-t_e)-\frac{\hat k \cdot \hat n_{re} |\vec v_\mathrm{DM}|}{c^2}\left|\vec x_r(t_r)-\vec x_e(t_e)\right|\right) \, \\
    &=\frac{\omega_\phi \left|\vec x_r(t_r)-\vec x_e(t_e)\right|}{c}\left(1 - \frac{\hat k \cdot \hat n_{re} |\vec v_\mathrm{DM}|}{c}\right) + \mathcal O(\mathrm{Shapiro}) \, ,
\end{align}
\end{subequations}
where we neglect the impact of the Shapiro delay on the signal.
The expression of $\alpha^\mathrm{DM}_{re}$ is therefore given by
\begin{equation}\label{eq:alpha_DM}
    \alpha^\mathrm{DM}_{re}(t_r) =\frac{\omega_\phi \left|\vec x_r(t_r)-\vec x_e(t_e)\right|}{2c}\left(1 - \frac{\hat k \cdot \hat n_{re} |\vec v_\mathrm{DM}|}{c}\right) \, .
\end{equation}

Let us now decompose the one-link response function into a slowly evolving part and a fast oscillating part. Let us  consider the observation time baseline $T_\mathrm{obs}$ such that the corresponding Fourier frequencies  are given by integer times $1/T_\mathrm{obs}$. Let us consider the Fourier frequency $q_0$ which is the closest to $f_\phi$, i.e.
\begin{equation}\label{eq:q}
	q_0= \mathrm{round}\left[f_\phi T_\mathrm{obs}\right]/T_\mathrm{obs} \, 
\end{equation}
where $\mathrm{round }\left[x\right]$ provides the closest integer to $x$. One can rewrite Eq.~(\ref{eq:Doppler_DM}) as
\begin{subequations}
\begin{equation}
    y^\mathrm{DM}_{re}(t_r) = \Re\left[\mathcal Y^\mathrm{DM}_{re}(t_r) e^{2\pi i q_0 t_r}\right] \, ,
\end{equation}
where 
\begin{equation}\label{eq:mathcal_Y_DM}
    \mathcal Y^\mathrm{DM}_{re} = 2i\left(\hat n_\mathrm{re} \cdot \hat e_v\right)\frac{\sqrt{8 \pi G \rho_\mathrm{DM}}v_\mathrm{DM}Q_d}{\omega_\phi c^2}\sin(\alpha^\mathrm{DM}_{re})e^{-i\beta^\mathrm{DM}_{re}}\, ,
\end{equation}
with
\begin{align}
    \beta^\mathrm{DM}_{re}&=\alpha^\mathrm{DM}_{re}+2\pi q_0 t_r - \phi_r  \,\nonumber\\
    &= \alpha^\mathrm{DM}_{re} - \Phi + 2 \pi (q_0-f_\phi) t_r  + \frac{2\pi f_\phi}{c}\frac{|\vec v_\mathrm{DM}|}{c}\hat k\cdot \vec x_r(t_r)  \, . \label{eq:beta_DM}
\end{align}
\end{subequations}

We now quickly show how to express TDI combinations in this framework. We will focus on the case of constant and equal arm-length $L$. In TDI, we encounter quantities such as
\begin{align}
    y_{ij}\left(t-\frac{L}{c}\right) = & Y_{ij}\left(t-\frac{L}{c}\right) e^{i\phi_i\left(t-\frac{L}{c}\right)}\, .
\end{align}
The quantity $Y_{ij}$ are slowly evolving with time such that $Y_{ij}(t-L/c)\approx Y_{ij}(t)$. For DM, this assumption holds since the amplitude slowly evolves with time (it changes with timescale $\tau(f)$ which is slow compared to the light travel time in the frequencies of interest) and if the spacecraft velocity is small compared to the speed of light. Then, we can write
\begin{subequations}
\begin{align}
    &\phi_i\left(t-\frac{L}{c}\right)=\omega_\phi t -\frac{\omega_\phi L}{c} -\vec k_\phi \cdot \vec x_i\left(t-\frac{L}{c}\right) +\Phi\, \\
    & =\omega_\phi t -\frac{\omega_\phi L}{c} -\vec k_\phi \cdot \vec x_i(t) + \Phi + \mathcal{O}(v_j/c \times L/x_i) \,\\
    &\approx \phi_i(t) - \frac{\omega_\phi L}{c} \, ,
\end{align} 
\end{subequations}
where $v_j$ is the velocity of the emitting spacecraft. This leads to
\begin{subequations}
\begin{align}
    y^\mathrm{DM}_{ij}\left(t-\frac{L}{c}\right) &\approx \Re\left[Y^\mathrm{DM}_{ij}(t) e^{i\phi_i(t)}e^{-i\omega_\phi L/c}\right] \,\\
    &=\Re\left[\mathcal Y^\mathrm{DM}_{ij}(t)e^{-i\omega_\phi L/c} e^{2\pi i q_0 t}\right] \, .
\end{align}
\end{subequations}
Applying recursively this relationship leads to
\begin{equation}
    D_{ijk}y^\mathrm{DM}_{kl}(t) = \Re\left[\mathcal Y^\mathrm{DM}_{kl}(t)e^{-2i\omega_\phi L/c} e^{2\pi i q_0 t}\right] \, ,
\end{equation}
and so on. Using this result, the first generation TDI combination can be written as 
\begin{subequations}
\begin{equation}
    X^\mathrm{DM}_1(t)=\Re\left[\mathcal X^\mathrm{DM}_1 (t) e^{2\pi i q_0 t}\right] \, ,
\end{equation}
and $\mathcal X^\mathrm{DM}_1$ is slowly evolving with time and is given by (using Eq.~\eqref{eq:TDI_X_gen}) 
\begin{align}
\mathcal X^\mathrm{DM}_1 &= \left(1-e^{-2i\omega_\phi L/c} \right) \times \,\nonumber \\
&\left[\mathcal Y^\mathrm{DM}_{13} -\mathcal Y^\mathrm{DM}_{12} + e^{-i\omega_\phi L/c}\left(\mathcal Y^\mathrm{DM}_{31}-\mathcal Y^\mathrm{DM}_{21}\right)\right] \nonumber \\
 &= 2i \sin \left(\frac{\omega_\phi L}{c}\right)e^{-i\omega_\phi L/c} \times \,\nonumber \\
&\left[\mathcal Y^\mathrm{DM}_{13} -\mathcal Y^\mathrm{DM}_{12} + e^{-i\omega_\phi L/c}\left(\mathcal Y^\mathrm{DM}_{31}-\mathcal Y^\mathrm{DM}_{21}\right)\right]\, .\label{eq:X1_sl}
\end{align}
\end{subequations}
Similarly the second generation TDI can be written as 
\begin{subequations}
\begin{equation}
    X^\mathrm{DM}_2(t)=\Re\left[\mathcal X^\mathrm{DM}_2 (t) e^{2\pi i q_0 t}\right]\, ,
\end{equation}
with the slowly evolving part
\begin{align}
\mathcal X^\mathrm{DM}_2 &= \left(1-e^{-2i\omega_\phi L/c}-e^{-4i\omega_\phi L/c}+e^{-6i\omega_\phi L/c}\right) \times \,\nonumber \\
&\left[\mathcal Y^\mathrm{DM}_{13} -\mathcal Y^\mathrm{DM}_{12} + e^{-i\omega_\phi L/c}\left(\mathcal Y^\mathrm{DM}_{31}-\mathcal Y^\mathrm{DM}_{21}\right)\right]\, .
\end{align}
Since $1-x-x^2+x^3=(1+x)(1-x)^2$, the previous expression becomes simply
\begin{align}
     \mathcal X^\mathrm{DM}_2 &=\left(1+e^{-2i\omega_\phi L/c}\right)\left(1-e^{-2i\omega_\phi L/c}\right)^2\times \,\nonumber \\
&\left[\mathcal Y^\mathrm{DM}_{13} -\mathcal Y^\mathrm{DM}_{12} + e^{-i\omega_\phi L/c}\left(\mathcal Y^\mathrm{DM}_{31}-\mathcal Y^\mathrm{DM}_{21}\right)\right] \,\nonumber\\
    &= \left(1+e^{-2i\omega_\phi L/c}\right)\left(1-e^{-2i\omega_\phi L/c}\right) \mathcal X^\mathrm{DM}_1 \,\nonumber\\
    &= 2i \sin \left(\frac{2\omega_\phi L}{c}\right) e^{-2i\omega_\phi L/c}\mathcal X^\mathrm{DM}_1\, . \label{eq:X2_sl}
\end{align}
\end{subequations}
We have now expressed the second generation TDI combination in time domain induced by scalar DM as function of a slowly evolving part (the $\mathcal{X}$ factors) and a fast oscillating part, oscillating at the closest Fourier bin to the wave Compton frequency.
This formulation is very convenient because in Fourier domain, the signal is simply the product of the Fourier transform of $\mathcal{X}$, which is numerically fast to obtain, and the Fourier transform of the $\exp(2\pi i q_0 t)$, which has an analytical solution (see \cite{Cornish07}).

\bibliographystyle{apsrev4-1}
\bibliography{main}
\end{document}